\documentclass[12pt]{article}
\usepackage{graphics}
\usepackage{enumerate}
\usepackage{amssymb,epsfig,amsmath,euscript,array}
\usepackage{axodraw}
\usepackage{subfigure}
\usepackage{color}
\makeatletter
\@addtoreset{equation}{section}
\makeatother

\definecolor{blue}{rgb}{0,0,1}

\definecolor{red}{rgb}{1,0,0}

\definecolor{green}{rgb}{0,1,0}

\def\s0#1#2{\mbox{\small{$ \frac{#1}{#2} $}}}
\def\0#1#2{\frac{#1}{#2}}

\newcounter{multieqs}

\newcommand{\be}{\begin{equation}}
\newcommand{\ee}{\end{equation}}
\newcommand{\eq}[1]{(\ref{#1})}

\newcommand{\bm}[1]{\mbox{\boldmath $#1$}}

\newcommand{\ks}{\mathbf{k'}}

\newcommand{\fssd}[1]{#1\!\!\!\!/}
\newcommand{\tr}{\,\text{tr}}

\newcommand{\Nc}{{N}_{\textrm{c}}}
\newcommand{\Nf}{{N}_{\textrm{f}}}

\def\bd{\begin{document}}
\def\ed{\end{document}}
\def\nn{\nonumber}
\def\bea{\begin{eqnarray}}
\def\eea{\end{eqnarray}}
\let\bm=\bibitem
\let\la=\label

\newcommand{\EQ}[1]{\begin{equation} #1 \end{equation}}
\newcommand{\AL}[1]{\begin{subequations}\begin{align} #1 \end{align}\end{subequations}}
\newcommand{\SP}[1]{\begin{equation}\begin{split} #1 \end{split}\end{equation}}
\newcommand{\ALAT}[2]{\begin{subequations}\begin{alignat}{#1} #2 \end{alignat}\end{subequations}}
\def\beqa{\begin{eqnarray}}
\def\eeqa{\end{eqnarray}}
\def\beq{\begin{equation}}
\def\eeq{\end{equation}}

\def\hf{{\textstyle \frac{1}{2}}}
\def\wbar{\bar w}
\def\mubar{\bar\mu}
\def\abar{\bar a}
\def\sigmabar{\bar\sigma}
\def\etabar{\bar\eta}
\def\zetabar{\bar\zeta}
\def\mubar{\bar\mu}
\def\nubar{\bar\nu}
\def\N{{\cal N}}
\def\sst{\scriptscriptstyle}
\def\thetabar{\bar\theta}
\def\one{\mbox{1 \kern-.59em {\rm l}}}
 \def\Nh{\hat{N}}

\newlength{\myVSpace}% the height of the box
\newcommand\xstrut{\raisebox{-.5\myVSpace}% symmetric behavior,
  {\rule{0pt}{\myVSpace}}%
}

\def\a{\alpha}      \def\da{{\dot\alpha}}
\def\b{\beta}       \def\db{{\dot\beta}}
\def\c{\gamma}  \def\G{\Gamma}  \def\cdt{\dot\gamma}
\def\d{\delta}  \def\D{\Delta}  \def\ddt{\dot\delta}
\def\e{\epsilon}        \def\vare{\varepsilon}
\def\f{\phi}    \def\F{\Phi}    \def\vvf{\f}
\def\h{\eta}
\def\k{\kappa}
\def\l{\lambda} \def\L{\Lambda}
\def\m{\mu} \def\n{\nu}
\def\o{\omega}
\def\p{\pi} \def\P{\Pi}
\def\r{\rho}
\def\t{\tau}
\def\th{\theta} \def\Th{\Theta} \def\vth{\vartheta}
\def\X{\Xeta}
\def\z{\zeta}

\def\cA{{\cal A}} \def\cB{{\cal B}} \def\cC{{\cal C}}
\def\cD{{\cal D}} \def\cE{{\cal E}} \def\cF{{\cal F}}
\def\cG{{\cal G}} \def\cH{{\cal H}} \def\cI{{\cal I}}
\def\cJ{{\cal J}} \def\cK{{\cal K}} \def\cL{{\cal L}}
\def\cM{{\cal M}} \def\cN{{\cal N}} \def\cO{{\cal O}}
\def\cP{{\cal P}} \def\cQ{{\cal Q}} \def\cR{{\cal R}}
\def\cS{{\cal S}} \def\cT{{\cal T}} \def\cU{{\cal U}}
\def\cV{{\cal V}} \def\cW{{\cal W}} \def\cX{{\cal X}}
\def\cY{{\cal Y}} \def\cZ{{\cal Z}}

\def\ua{\underline{\alpha}}
\def\ub{\underline{\phantom{\alpha}}\!\!\!\beta}
\def\uc{\underline{\phantom{\alpha}}\!\!\!\gamma}
\def\um{\underline{\mu}}
\def\ud{\underline\delta}
\def\ue{\underline\epsilon}
\def\una{\underline a}\def\unA{\underline A}
\def\unb{\underline b}\def\unB{\underline B}
\def\unc{\underline c}\def\unC{\underline C}
\def\und{\underline d}\def\unD{\underline D}
\def\une{\underline e}\def\unE{\underline E}
\def\unf{\underline{\phantom{e}}\!\!\!\! f}\def\unF{\underline F}
\def\unm{\underline m}\def\unM{\underline M}
\def\unn{\underline n}\def\unN{\underline N}
\def\unp{\underline{\phantom{a}}\!\!\! p}\def\unP{\underline P}
\def\unq{\underline{\phantom{a}}\!\!\! q}
\def\unQ{\underline{\phantom{A}}\!\!\!\! Q}
\def\unH{\underline{H}}

\def\As {{A \hspace{-6.4pt} \slash}\;}
\def\bs {{b \hspace{-6.4pt} \slash}\;}
\def\Ds {{D \hspace{-6.4pt} \slash}\;}
\def\ds {{\del \hspace{-6.4pt} \slash}\;}
\def\ss {{\s \hspace{-6.4pt} \slash}\;}
\def\ks {{ k \hspace{-6.4pt} \slash}\;}
\def\ps {{p \hspace{-6.4pt} \slash}\;}
\def\pas {{{p_1} \hspace{-6.4pt} \slash}\;}
\def\pbs {{{p_2} \hspace{-6.4pt} \slash}\;}

\def\Fh{\hat{F}}
\def\Vh{\hat{V}}
\def\Xh{\hat{X}}
\def\ah{\hat{a}}
\def\xh{\hat{x}}
\def\yh{\hat{y}}
\def\ph{\hat{p}}
\def\xih{\hat{\xi}}

\def\psit{\tilde{\psi}}
\def\Psit{\tilde{\Psi}}
\def\tht{\tilde{\th}}

\def\At{\tilde{A}}
\def\Qt{\tilde{Q}}
\def\Rt{\tilde{R}}
\def\Nt{\tilde{N}}

\def\at{\tilde{a}}
\def\st{\tilde{s}}
\def\ft{\tilde{f}}
\def\pt{\tilde{p}}
\def\qt{\tilde{q}}
\def\vt{\tilde{v}}
\def\nt{\tilde{n}}

\def\delb{\bar{\partial}}
\def\bz{\bar{z}}
\def\bD{\bar{D}}
\def\bB{\bar{B}}

\def\bk{{\bf k}}
\def\bl{{\bf l}}
\def\bp{{\bf p}}
\def\bq{{\bf q}}
\def\br{{\bf r}}
\def\bx{{\bf x}}
\def\by{{\bf y}}
\def\bR{{\bf R}}
\def\bV{{\bf V}}

\def\d{\delta}\def\D{\Delta}\def\ddt{\dot\delta}

\def\pa{\partial} \def\del{\partial}
\def\xx{\times}
\def\uno{\mbox{1 \kern-.59em {\rm l}}}

\def\trp{^{\top}}
\def\inv{^{-1}}
\def\dag{{^{\dagger}}}
\def\pr{^{\prime}}

\def\rar{\rightarrow}
\def\lar{\leftarrow}
\def\lrar{\leftrightarrow}

\def\one{1\!\!1\,\,}
\def\im{\imath}
\def\jm{\jmath}

\newcommand{\slsh}[1]{/ \!\!\!\! #1}

\def\vac{|0\rangle}
\def\lvac{\langle 0|}

\def\hlf{\frac{1}{2}}
\def\ove#1{\frac{1}{#1}}

\def\Box{\square}
\def\ZZ{\mathbb{Z}}
\def\CC#1{({\bf #1})}
\def\bcomment#1{}
\def\bfhat#1{{\bf \hat{#1}}}
\def\VEV#1{\left\langle #1\right\rangle}
\def\vev#1{\langle{#1}\rangle}

\newcommand{\ex}[1]{{\rm e}^{#1}} \def\ii{{\rm i}}

\def\rr{{\rm r}} \def\rs{{\rm s}}\def\rv{{\rm v}}
\def\ri{{\rm i}}\def\rj{{\rm j}}

\newcommand{\lrbrk}[1]{\left(#1\right)}
\newcommand{\sfrac}[2]{{\textstyle\frac{#1}{#2}}}

\font\mybb=msbm10 at 12pt
\def\bb#1{\hbox{\mybb#1}}

\font\myBB=msbm10 at 18pt
\def\BB#1{\hbox{\myBB#1}}

\begin{document}
\noindent
%\hspace*{13cm}DESY 06-122\\
%\hspace*{13cm}HD-THEP 06-17\\
\hfill DESY 06-122\\
$\text{}$\hfill HD-THEP 06-17\\

\vspace{25pt}

\begin{center}

{\Large \bf Do Instantons Like a Colorful Background?\\[1.5ex]
}

\vspace*{30pt}

{\bf Holger Gies$^1$, Joerg Jaeckel$^2$, Jan
  M.~Pawlowski$^1$ and Christof Wetterich$^1$}

{\small \em
  {}$^1$Institut fuer Theoretische Physik, Philosophenweg 16, D-69120 Heidelberg, Germany\\
  {}$^2$Deutsches Elektronen-Synchrotron DESY,
  Notkestrasse 85, D-22607  Hamburg, Germany\\

\vspace{10pt}

{\sffamily \tt
H.Gies@thphys.uni-heidelberg.de, joerg.jaeckel@desy.de, J.Pawlowski@thphys.uni-heidelberg.de, C.Wetterich@thphys.uni-heidelberg.de}
}

\vspace{30pt}
{\bf Abstract}
\end{center}
We investigate chiral symmetry breaking and color symmetry breaking in
QCD. The effective potential of the corresponding scalar condensates
is discussed in the presence of non-perturbative contributions from
the semiclassical one-instanton sector. We concentrate on a
color singlet scalar background which can describe chiral
condensation, as well as a color octet scalar background which can
generate mass for the gluons.
Whereas a non-vanishing singlet chiral field is favored by the
instantons, we have found no indication for a preference of color
octet backgrounds.
\noindent {} \setcounter{page}{0} \thispagestyle{empty}

\newpage

\section{Introduction}
Instantons, being pseudo-particles associated with tunneling
processes, generate genuine non-perturbative effects in QCD.  In the
seminal work of 't Hooft \cite{'tHooft:fv} it was realized that they
mediate an effective interaction between (light) quarks
\cite{'tHooft:fv,Shifman:uw,Callan:1977gz,Shuryak:1982hk,Schafer:1996wv,%
Pawlowski:1996ch,Wetterich:2000ky,Meggiolaro:2000kp}.  This
``instanton interaction'' is attractive in the color singlet channel;
hence, instantons presumably play a role in the mechanism of chiral
symmetry breaking \cite{Diakonov:vw,Diakonov:1985eg,Carter:1999xb}. In
addition, they also provide for an interaction in color octet channels
or in color triplet and sextet ``diquark'' channels. Mean-field
computations based on a point-like instanton interaction have been
employed as a central tool for investigations of color
superconductivity at high baryon density
\cite{Bailin:bm,Alford:1997zt,Rapp:1997zu,Alford:1998mk,Berges:1998rc},
or for a description the baryon and meson spectrum and interactions in the
vacuum in a Higgs picture with spontaneous color symmetry breaking
\cite{Wetterich:1999vd,Wetterich:2000ky,Wetterich:2000pp}. This
phenomenologically quite successful scenario requires a
quark--anti-quark condensate in the color octet channel, giving rise
to the question as to whether instantons support quark condensation in
this channel. 

Symmetry breaking by a condensation phenomenon requires an interaction
that lowers the free energy if condensates are formed. The bosonic
condensates can be quark bilinears or even higher-order composites. In
the case of instantons, a rich interaction structure is indeed
provided: for $\Nf$ light quarks, instantons typically induce an
interaction between $2\Nf$ quark fields, which can be paired in many
ways. This is one of the reasons why instanton-induced multi-fermion
interactions have often been used as a starting point for
investigations in the mean-field approximation. However, in the
approximation of a point-like multi-fermion interaction, mean-field
theory is ambiguous: by means of a Fierz transformation, the quarks
can be grouped in different ways. For example, products involving
color non-singlet Lorentz scalars can be exchanged by products of
color singlets in vector or tensor representations of the Lorentz
group and vice versa. In view of this ambiguity, the relative strength
between color octet and singlet channels remains undetermined, since
the color octet channels can be completely removed or enhanced by
suitable re-orderings \cite{Jaeckel:2003xy}. Similar problems arise for
the other colored channels used in the high-density computations.

For further progress towards reliable computations, the Fierz
ambiguity of the mean-field computation has to be resolved. This can be
done in different ways. A first possibility explicitly includes the
fluctuations of composite bosons after partial bosonization. Then, the
dependence on the particular choice of bosonization (Fierz ambiguity)
gets substantially reduced, as demonstrated by functional
renormalization group techniques \cite{Jaeckel:2002rm}. A second
approach attempts to resolve the ambiguity by explicitly taking the
momentum dependence of the instanton-induced vertex into
account. Finally, we propose a third method in this article that
avoids altogether the use of the multi-fermion vertex and rather
computes directly the instanton contribution to the free energy in the
presence of selected condensates. The various approaches have
different strengths and shortcomings, and a reliable picture will
probably only emerge by a combination of them. 

The advantage of a study of the momentum dependence of the
instanton-induced vertex is based on the observation that pole-like
structures which arise from the effective exchange of
quark--anti-quark or quark--quark bound states can be associated to
the given channel of the bound state. In contrast to a point-like
interaction, such pole structures can no longer be moved to another
channel by Fierz reordering.  Momentum-dependent vertex functions can
be dealt with using functional methods, such as Dyson-Schwinger
equations, $N$PI effective actions, functional renormalization group
(RG) or suitable combinations.  In particular, we envisage the
functional RG as a promising approach towards a quantitative study for
the condensation phenomena at hand, for reviews see
\cite{Litim:1998nf,Berges:2000ew,Polonyi:2001se,Pawlowski:2005xe}.
The computation of the flow equations involves only a narrow momentum
range around a given renormalization scale $k$, thus reducing the
impact of an incomplete knowledge of the detailed momentum dependence
of the full propagators and vertices; see, e.g.,
\cite{Ellwanger:1995qf,Pawlowski:2003hq,Fischer:2004uk,Pawlowski:2005xe,Blaizot:2005xy}.
Its application to the present problem requires an implementation for
the non-perturbative sector of gauge theories, e.g.\
\cite{Litim:1998nf,Reuter:1993kw,Ellwanger:1995qf,Freire:2000bq,Gies:2002af,Pawlowski:2003hq,Fischer:2004uk},
also employing bosonization techniques as developed in
\cite{Gies:2001nw,Gies:2002hq,Jaeckel:2002rm,Gies:2005as,Pawlowski:2005xe},
or NPI- and NPPI-flows as discussed in
\cite{Polonyi:2001se,Wetterich:2002ky,Jaeckel:2003uz,Pawlowski:2005xe}.
In particular, the Fierz-type ambiguity of the present problem can be
resolved within a 2PPI-effective-action approach, since all possible
(local) fermionic pairings are effectively taken into account by this
approach \cite{Jaeckel:2003xy}.

In this work, we consider a more direct approach to instanton-induced
color symmetry breaking by taking advantage of the following
observation: possible condensates can be viewed as background fields
that are coupled to quarks and gluons via Yukawa and gauge
interactions. We concentrate here on scalar color singlet and octet
condensates. In presence of a singlet condensate, all three light
quarks become massive, thus influencing the weight of the fermion
determinant in the instanton calculation. Additional octet condensates
induce a mass split between an octet of fermions (here associated with the
baryon octet) and a singlet. Furthermore, all gluons acquire mass
through the Higgs mechanism. Both effects modify the instanton
contribution to the free energy. In particular, the effective
condensate-dependent gluon mass acts as an effective infrared cutoff,
strongly suppressing the contribution of instantons of size larger
than the inverse gluon mass. Furthermore, the infrared cutoff stops
the running of the gauge coupling such that the gauge coupling remains
small for sufficiently large octet condensates, and perturbation
theory becomes applicable. By computing the instanton contribution to
the free energy in presence of the condensates, we get access to those
parts of the effective potential that violate the axial
U(1)${}_{\mathrm{A}}$ symmetry. Under the hypothesis that these parts
dominate the octet dependence of the potential, we may try to draw
conclusions if the minimum occurs for vanishing or non-vanishing octet
condensate. Our computation of this response is based on two
theoretical concepts: on the one hand, the full functional integral is
evaluated in the semiclassical one-instanton approximation. On the
other hand, the decoupling of massive modes is taken care of by a
proper threshold behavior of the running coupling, as it is suggested
by the functional RG.

Our method needs assumptions how the non-vanishing singlet and octet
condensates influence the masses of quarks and gluons. In practice,
this is done by an ansatz for the effective action which describes the
couplings of quarks and gluons to the color singlet and octet
condensates. Apart from the restrictions imposed by color and flavor
symmetry, the details of this effective action are not known. This is
one of the most severe restrictions on the quantitative reliability of
our computation. Nevertheless, the qualitative features of mass
generation for quarks and gluons can be captured in a simple
picture. We consider here a local interaction with low powers of the
condensates, in particular the chiral color singlet scalar
$\sigma_{ab}$ with flavor indices $a,b,\dots$ and the color octet
scalar $\chi_{ab,ij}$ with non-trivial structure for flavor and color
($i,j,\dots$). Our ansatz for the interactions between the condensate
fields, quarks and gluons can be summarized in the following Euclidean
effective Lagrangian \cite{Wetterich:2000pp},
\begin{eqnarray}
\label{model}
\mathcal{L}&=&i Z_{\psi}\bar{\psi}_{i}\fssd{D}_{ij}\,\psi_{j}
+\frac{1}{2}\,F^{\mu\nu}_{ij} F_{ji,\mu\nu}
\\\nonumber
&&+Z_{\chi}\tr\,\{(D^{\mu}\chi)_{ij}^{\dagger}(D^{\mu}\chi)_{ij}\}
+Z_{\sigma}\tr\,\{\partial^{\mu}\sigma^{\dagger}\partial^{\mu}\sigma\}
\\\nonumber
&&-i
Z_{\psi}\bar{\psi}_{i}\big[(h\sigma\delta_{ij}+\tilde{h}\chi_{ij})
\frac{1+\gamma^{5}}{2}+(h\sigma^{\dagger}\delta_{ij}+
\tilde{h}\chi^{\dagger}_{ji})
\frac{1-\gamma^{5}}{2}\big]\psi_{j}
\\\nonumber
&&+U_{0}(\sigma,\chi).
\end{eqnarray}
Here, we have included all power-counting relevant and marginal
interaction operators as well as an effective potential for the
background fields for completeness. In Eq.~\eqref{model}, we treat
$\sigma$ and $\chi_{ij}$ as $3\times3$ matrices in flavor space and
contract over the flavor indices of the quarks.  A successful
phenomenology of QCD based on an effective Lagrangian of this form has
been worked out in
\cite{Wetterich:1999vd,Wetterich:2000pp,Wetterich:1999sh}. 

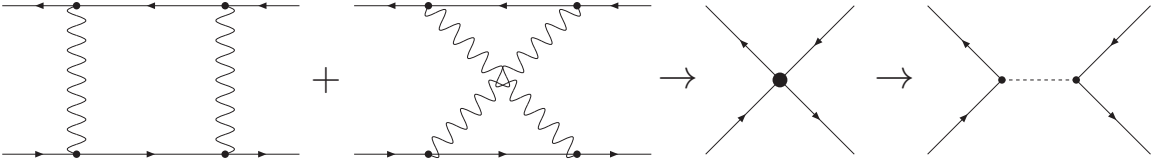
\begin{figure}[t]
\begin{center}
\subfigure{}{\scalebox{0.7}[0.7]{%\fbox{
\begin{picture}(190,140)(-10,0)
  \SetOffset(3,10)
  \ArrowLine(40,100)(0,100)
  \ArrowLine(120,100)(40,100)
  \ArrowLine(160,100)(120,100)
  \ArrowLine(0,20)(40,20)
  \ArrowLine(40,20)(120,20)
  \ArrowLine(120,20)(160,20)
  \Vertex(40,100){2}
  \Vertex(120,100){2}
  \Vertex(40,20){2}
  \Vertex(120,20){2}
  \Photon(40,100)(40,20){-5}{7.5}
  \Photon(120,100)(120,20){5}{7.5}
  \Text(175,60)[c]{\scalebox{1.8}[1.8]{$+$}}
\end{picture}}}%}
%\hspace{1cm}
\subfigure{}{\scalebox{0.7}[0.7]{%\fbox{
\begin{picture}(190,140)(-10,0)
  \SetOffset(3,10)
  \ArrowLine(40,100)(0,100)
  \ArrowLine(120,100)(40,100)
  \ArrowLine(160,100)(120,100)
  \ArrowLine(0,20)(40,20)
  \ArrowLine(40,20)(120,20)
  \ArrowLine(120,20)(160,20)
  \Vertex(40,100){2}
  \Vertex(120,100){2}
  \Vertex(40,20){2}
  \Vertex(120,20){2}
  \Photon(40,100)(120,20){-5}{10.5}
  \Photon(120,100)(40,20){5}{10.5}
  \Text(175,60)[c]{\scalebox{1.8}[1.8]{$\rightarrow$}}
\end{picture}}}%}
\subfigure{}{\scalebox{0.7}[0.7]{%\fbox{
\begin{picture}(120,140)(-10,0)
  \SetOffset(3,10)
  \ArrowLine(40,60)(0,100)
  \ArrowLine(0,20)(40,60)
  \ArrowLine(80,100)(40,60)
  \ArrowLine(40,60)(80,20)
  \Vertex(40,60){4}
  \Text(102,60)[c]{\scalebox{1.8}[1.8]{$\rightarrow$}}
\end{picture}}}%}
\subfigure{}{\scalebox{0.7}[0.7]{%\fbox{
\begin{picture}(190,140)(-10,0)
  \SetOffset(3,10)
  \ArrowLine(40,60)(0,100)
  \ArrowLine(0,20)(40,60)
  \DashLine(40,60)(80,60){2}
  \ArrowLine(120,100)(80,60)
  \ArrowLine(80,60)(120,20)
  \Vertex(40,60){2}
  \Vertex(80,60){2}
\end{picture}}}%}
\end{center}
\vspace{-1.0cm}
\caption{Box diagrams with fundamental QCD interactions (left two
  diagrams) generate effective (nonlocal) four fermion interactions
  (middle). Using rebosonization \cite{Gies:2001nw} these can be
  translated to (approximately local) Yukawa interactions interactions
  with propagating composite bosons (right).}
\label{box}
\end{figure}

 Obviously, the reliability of our conclusions will depend on whether
the ansatz \eqref{model} gives a qualitatively correct picture for the
response to non-vanishing condensates. We therefore present a few
additional arguments for its motivation. Associating the condensates
$\sigma,\chi$ with corresponding fermion composites $\sim \bar\psi
\psi$, the interactions of the type specified in Eq.~\eqref{model}
arise naturally from fundamental QCD, as can be studied with
techniques developed in
\cite{Gies:2001nw,Gies:2002hq,Jaeckel:2002rm,Jaeckel:2003uz,Gies:2005as}.
In particular, box diagrams of the type shown in Fig.~\ref{box} play
an important role.  In this work, we choose the viewpoint that these
effective interactions are present in the dominant momentum region for
the instanton contribution, being generated by
U(1)${}_{\text{A}}$-preserving interactions, also partly at higher momentum
scales. We do not attempt here to compute the parameters appearing in
the effective action \eqref{model} except for the gauge coupling. For
a qualitative study, we treat the Yukawa couplings $h,\tilde h$ as
well as the wave function renormalization factors $Z_\psi$,
$Z_\sigma$, $Z_\chi$ as free parameters.

In presence of non-vanishing background fields $\sigma$ chiral
symmetry is broken, whereas $\chi$ acts like the Higgs scalar, giving
masses to gluons and quarks. For the present purpose, it suffices to
investigate in detail the following two directions in field space:
\begin{equation}
\label{condensates}
\sigma_{ab}=\sigma\delta_{ab},\qquad
\chi_{ab,ij}=\frac{1}{\sqrt{6}}\chi(\delta_{ia}\delta_{jb}
-\frac{1}{3}\delta_{ij}\delta_{ab}).
\end{equation}
These configurations correspond to the condensates of standard chiral
symmetry breaking and a color-flavor locked \cite{Schafer:1998ef}
combination of quarks and anti-quarks, respectively.
In this background, all fermions aquire mass,
\begin{equation}
  \label{fermionmasses}
M_{1}=h\sigma+\frac{8}{3\sqrt{6}}\tilde{h}\chi,\,\,\, \qquad
M_{8}=h\sigma-\frac{1}{3\sqrt{6}}\tilde{h}\chi,
\end{equation}
with a split between the octet mass $M_8$ and the singlet mass $M_1$
for $\chi\neq0$. The fermion determinant in the instanton contribution
depends only on $M_8$ and $M_1$. We use the freedom of scaling of the
fields $\sigma$ and $\chi$ to set $h=\tilde h=1$. In this
normalization, $\sigma$ and $\chi$ are directly related to the
masses. In the Higgs picture of the QCD vacuum, the expectation value
for $M_8$ should be associated with the mass of the lowest baryon
octet and $M_1$ with a baryon singlet, possibly $\Lambda(1405)$,
yielding \cite{35A} $M_8=1.15$GeV, $M_1=-1.4$GeV or $\sigma_0=866$MeV,
$\chi_0=-2.08$GeV. In our approach, we treat $\sigma$ and $\chi$ as
free variables. The octet condensate in \eqref{fermionmasses} provides
for an equal mass for all eight gluons,
\begin{equation}
\label{gluonmass}
M_{g}=Z_{\chi}^{1/2}g|\chi|.
\end{equation}
Here, $g$ is the renormalized coupling taken at an appropriate
scale. The $\chi$ dependence of $g$ will be discussed in detail
below. Then $Z_\chi$ remains the only undetermined parameter of our
ansatz. The phenomenological ansatz of \cite{Wetterich:2000pp,35A}
associates $M_{g}$ with the average mass of the lowest spin-one meson
octet, $M_g\simeq 850$MeV and suggests $Z_\chi^{1/2}\simeq1/15$. 

The paper is organized as follows. In Sect. \ref{sec-instanton}, we
discuss the various effects of the quark and gluon masses on the
instanton integral. In the subsequent Sect. \ref{asymptotic}, we
discuss the asymptotic behavior of the instanton contribution to the
free energy. In Sect. \ref{solution}, we investigate which condensate
backgrounds are preferred by the instantons. Our conclusions are
presented in Sect.~\ref{sec:conc}.

\section
{Effective potential in one instanton approximation}\label{sec-instanton}

Consider a given background of scalar fields $\sigma$ and $\chi$, as
introduced above. Our aim is to compute the instanton contribution to
the effective potential for $\sigma$ and $\chi$ in the presence of
fluctuating quarks and gluons.  For homogeneous $\sigma$ and $\chi$,
the effective action $\Gamma$ thus decomposes into 
\begin{equation}
\Gamma[\sigma,\chi]\equiv\Omega\, U(\sigma,\chi)=\Omega \, \big(
U_0(\sigma,\chi)+U_{\text{inst}}(\sigma,\chi)+U_{\text{inst}}^\ast(\sigma,\chi)
\big), \label{effact} 
\end{equation}
where $\Omega$ denotes the spacetime volume. The non-anomalous
contribution $U_0(\sigma,\chi)$ conserves the axial
U(1)${}_{\text{A}}$ symmetry and will not be computed here. The
anomalous contribution $U_{\text{inst}}$ is induced by configurations
with non-trivial topology, mediating also $\textrm{U}(1)_\text{A}$
violation \cite{'tHooft:1986nc}. We determine this part in
semi-classical approximation based on instanton methods.  In
particular, we resort to the approximation of a gas of dilute
instantons in which $U_{\text{inst.}}$ can be expressed by an integral
over the instanton size $\rho$ and the product of gluonic
(incl. ghosts) and fermionic fluctuation determinants in a one-instanton
background (see Appendix~\ref{instgas}),
\begin{equation}
U_{\text{inst}}=-\frac{1}{\Omega} \, \int_0^\infty d \rho\,
 \exp(-8\pi^2/g^2(\rho))\, \Delta_{\text{gl}}(\chi,\rho)\, \det\,
 M_{\psi,ij}. 
\label{Uinst}
\end{equation}
Here, the exponential factor reflects the classical action of the
instanton, and $\Delta_{\text{gl}}$ summarizes the contributions from
gluons and ghosts in the instanton background. The last factor, with
\begin{equation}
M_{\psi,ij}=-\fssd{D}\,_{ij}+\sigma\delta_{ij}+\chi_{ij},
\end{equation}
represents the fermion determinant which is of central interest to our
work. In particular, it contains the zero modes of the Dirac operator
which are responsible for anomalous contributions and give rise to a
strong $\sigma$ and $\chi$ dependence even for small values of these
fields.

It is useful to decompose Eq.~(\ref{Uinst}) into a factor
$\zeta_{\text{z}}(\sigma,\chi,\rho)$ arising from the fermionic zero
modes, and another non-zero-mode factor
$\zeta_{\text{n}}(\sigma,\chi,\rho)$ that summarizes all remaining
(classical, gluonic, fermionic) contributions:
\begin{equation}
U_{\text{inst}}(\sigma,\chi)=-\int d\rho\,
\zeta_{\text{n}}(\sigma,\chi,\rho)\,
\zeta_{\text{z}}(\sigma,\chi,\rho). \label{Uinst2}
\end{equation}
All contributions have been studied frequently in the literature,
beginning with the seminal work of 't Hooft \cite{'tHooft:fv}. As
important new aspects, we include the color octet scalar and take the
threshold behavior due to decoupling of massive modes into account.

\subsection{Lowest order in the background fields}

Assuming that $\sigma$ and $\chi$ are small compared to all other
scales, their main influence arises from the zero-mode contribution.
In particular, the non-zero-mode factor $\zeta_{\textrm{n}}$ does not
depend on the scalar fields to lowest order.  For an SU($\Nc$) gauge
theory with $\Nf$ flavors, $\zeta_{\textrm{n}}$ reads\footnote{In 
  Appendix~\ref{assembling}, we briefly review the contributions from
  the zero and the non-zero modes starting from results given in
  \cite{'tHooft:fv,Bernard:1979qt}. Moreover, we use this appendix to
  introduce our regularization scheme.} \cite{'tHooft:fv},
\begin{eqnarray}
\label{nonzeromodes}
\zeta_{\textrm{n}}(\rho)
=
%\int d\rho
D_{\textrm{S}}\rho^{-5}
\left(\frac{8\pi^2}{g^{2}(\rho)}\right)^{2\Nc}\exp\left(
-\frac{8\pi^2}{g^{2}(\rho)}\right).
\end{eqnarray}
Here, $D_\textrm{S}$ is a scheme-dependent constant. A discussion of
the scheme dependence including the difference between massive and
massless regularization schemes can be found in
Appendix~\ref{rgscheme}. Our scheme has been motivated by the
functional RG which generically provides for mass-dependent schemes
that automatically account for a proper decoupling of massive modes.
This is a convenient feature of our RG-inspired scheme; however, we
observe no qualitative scheme dependencies of our results.  For
example, to zeroth order in the fields, $D_\textrm{S}$ in our RG
regularization scheme is given by (see
\cite{Luscher:1981zf,Hasenfratz:1981tw,'tHooft:1986nc} and
Appendix~\ref{assembling})
\begin{equation}
\label{constants}
D_{\overline{\textrm{RG}}}=D_{\overline{\textrm{MS}}}
=%s^{\beta_{0}}_{\textrm{I}}
\frac{2\exp(\frac{5}{6})}{\pi^2(\Nc-1)!(\Nc-2)!}
\exp(-1.51137\Nc+0.29175\Nf)=6.005\times10^{-3},
\end{equation}
where the last equality holds for $\Nc=\Nf=3$.

As discussed in the Appendices~\ref{rgscheme}, \ref{assembling}, our
$\overline{\textrm{RG}}$ scheme is constructed such that it matches
the $\overline{\textrm{MS}}$ scheme in the small mass limit.  It was
demonstrated in \cite{Ringwald:1999ze} that the
$\overline{\textrm{MS}}$ scheme gives satisfactory agreement with
lattice data in the ultraviolet. Without a color-flavor mixing mass
matrix $(\chi=0)$, the eigenmodes of $\fssd{D}\,$ are also eigenmodes
of $M_\psi$ and we are led to \cite{'tHooft:fv,Pawlowski:1996ch}
\begin{eqnarray}
\label{zeromodes}
\zeta_{\textrm{z}}(\rho,\sigma,\chi)=
\langle \langle \det{}_{\textrm{flavor}}
\langle\psi_0(a,i)|M_{\psi,ij}|\psi_0(b,j)\rangle\rangle_{\textrm{SU}(3)}=:
-\rho^{\Nf}V(\sigma,\chi),
\end{eqnarray}
where the inner angled brackets denote the scalar product of the zero
modes $\psi_0$, and the outer angled brackets denote a group average
over all possible directions for the instanton in color space. In the
last step, we have separated off the simple $\rho$ dependence $\sim
\rho^{\Nf}$ and defined the auxiliary potential $V(\sigma,\chi)$.  We
have also used the persistence of \mbox{(quasi-)zero} modes in the presence
of the regularization \cite{Pawlowski:1996ch}. For $\chi\neq 0$,
$\fssd{D}\,\,$ and $M_{\psi}$ do not commute in general, e.g.\ for
condensates $\chi$ with \eq{condensates}. The eigenmodes of
$\fssd{D}\,$ and $M_\psi$ do not agree anymore for $\chi\neq
0$. Therefore, strictly speaking, Eq.~\eqref{zeromodes} does not hold
in general.  However, in leading order of an expansion in $\chi$ and
$\sigma$ it holds true, as shown in Appendix~\ref{zerodet}.  Inserting
these findings into Eq.~(\ref{Uinst2}), we obtain
\begin{equation}
\label{potential}
U_{\textrm{inst}}(\sigma,\chi)=V(\sigma,\chi)\int d\rho\, \rho^{\Nf}
\zeta_{\textrm{n}}(\rho)=:\zeta\, V(\sigma,\chi).
\end{equation}
For $\Nf<4$, $\zeta$ is a finite number for physically admissible
  running couplings from the UV to the IR as discussed in
  Appendix~\ref{runcoup}. For small $\sigma$ and $\chi$, the potential
  $V(\sigma,\chi)$ carries all dependence on the scalar condensates.

So far our discussion has made no use of a specific color or flavor
structure for the background fields. Let us now specialize to the
condensates specified in Eq.~\eqref{condensates}. Using the
gauge-group averages computed in \cite{Shifman:uw} we find (see
Appendix~\ref{zerodet} for details)
\begin{equation}
V(\sigma,\chi)=-\sigma^{3}+\frac{1}{72}\sigma\chi^{2}+\frac{1}{648
\sqrt{6}}\chi^{3}=-(\sigma+
\frac{1}{6\sqrt{6}} \chi)^2(\sigma-\frac{1}{3\sqrt{6}} \chi).
\label{Vpot}
\end{equation}
In this crude approximation where $U(\sigma,\chi)=\zeta
V(\sigma,\chi)$ with $\zeta$ being a field-independent constant, we
observe two flat directions, $\sigma=-\frac{1}{6\sqrt{6}} \chi$ and
$\sigma=\frac{1}{3\sqrt{6}} \chi$, but no global minimum. In fact,
$V(\sigma,\chi)$ is unbounded from below, similar to the findings in
\cite{Jaeckel:2003xy}.  In the present case, this simply signals the
breakdown of the approximation of small $\sigma$ and $\chi$.

Let us assume for a moment that the potential becomes stable beyond
this approximation or by the inclusion of $U_0(\sigma,\chi)$ (cf.
Eq.~\eqref{effact}). Then one might speculate that the first flat
direction, $\sigma=-\frac{1}{6\sqrt{6}} \chi$, which is a line of
local minima for $\sigma<0$, characterizes a global minimum (the
second flat direction is not even a local minimum). However, in this
case, the ratio
$r=\left|\frac{\sigma}{\chi}\right|=\frac{1}{6\sqrt{6}}\approx 0.068$
is far from the phenomenologically reasonable range $r\sim0.4$
\cite{35A}.  Since $V$ is completely determined by the
zero modes of the massless Dirac operator, this flat direction will
not be lifted by the inclusion of higher order corrections in the
bosonic fields in the 1-instanton approximation, as long as the split
into zero- and non-zero-mode parts remains justified.  A similar flat
direction was also found in \cite{Jaeckel:2003xy}.

\begin{figure}[t]
\begin{center}
\scalebox{0.75}[0.75]{
\begin{picture}(190,140)(40,0)
\includegraphics[width=9.5cm]{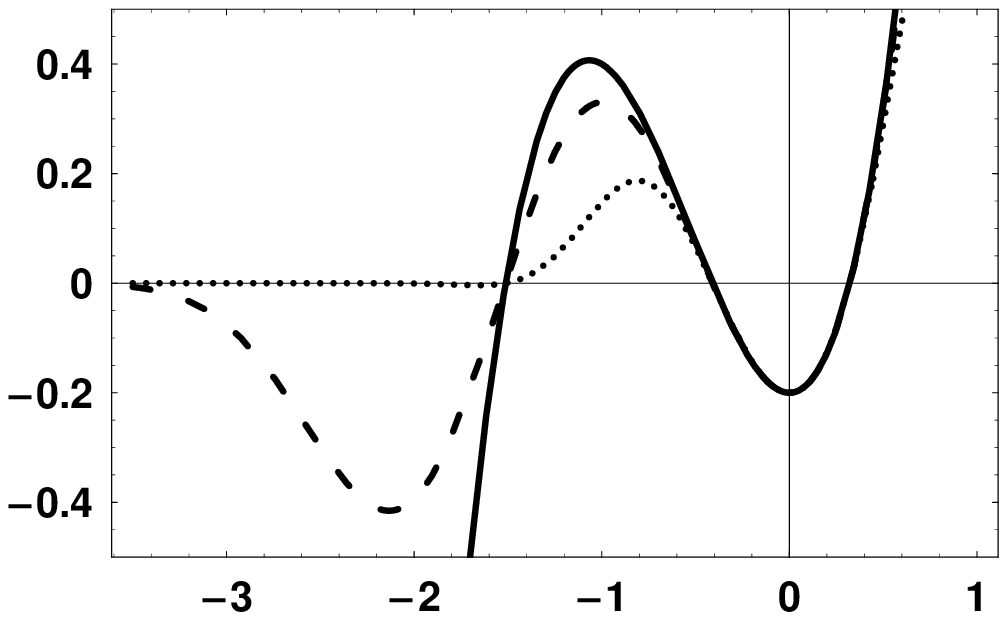}
\Text(-10,-10)[c]{\scalebox{1.6}[1.6]{$\chi$}}
\Text(-290,150)[c]{\scalebox{1.6}[1.6]{$V(\chi)$}}
\end{picture}
}
\end{center}
\vspace{-0.3cm}
\caption{Schematic plot of $V(\sigma,\chi)$ at fixed
  $\sigma>0$. Without higher-order corrections, the potential is
  unbounded from below (solid line) with a local minimum at $\chi=0$.
  If the cutoff mechanism provided by the higher-order corrections is
  strong (dotted line) the global minimum remains at $\chi=0$.
  However, if the suppression sets in only at rather large values of
  $\chi$ (dashed line) we have a global minimum at $\chi\neq 0$ in
  addition to a local one at $\chi=0$.}
\label{figsigmaguess}
\end{figure}

Let us furthermore assume that, for instance, $U_0$ induces a nonzero
VEV for $\sigma$. Since $V(\sigma,\chi)$ prefers a positive $\sigma$,
the resulting potential $V(\sigma,\chi)$ in the $\chi$ direction looks
like the solid line sketched in Fig.~\ref{figsigmaguess}. The case of
no color octet condensate, $\chi=0$, then is a local minimum. For
larger $\chi$, the higher-order corrections from the non-zero-mode
contribution and the threshold effects will set in, stabilizing the
potential in $\chi$ direction. Now it is a dynamical question as to
whether this stabilization sets in early, i.e., for rather small
$\chi$, such that no other minimum is induced (dotted line). Or
stabilization could only modify the region of large $\chi$ (dashed)
line, such that the $\chi^3$ term of Eq.~\eqref{Vpot} wins out in
between and induce a color octet condensate.

The second scenario of color octet condensate formation seems more
difficult to be realized, since the $\sim \sigma^3$ term and the $\sim
\sigma\chi^2$ are of opposite sign and the coefficient of the $\chi^3$
term is rather small. Unfortunately, the small coefficients in front
of $\sigma\chi^2$ and $\chi^3$ in the potential \eqref{Vpot} will
limit even the qualitative reliability of our investigation. As an
effect of the color averaging, the potential in the $\chi$ direction
is almost flat for a given value of $\sigma$, in contrast to the
pronounced potential in the $\sigma$ direction. For a given $\sigma$,
the weak dependence of $V$ on the ``direction'' $\chi/\sigma$ could
easily be overwhelmed by corrections in higher orders in $\sigma$ and
$\chi$ that are much more difficult to control. Despite this caveat, a
quantitative analysis remains interesting and will be presented in the
next sections.

Let us close this lowest-order consideration with the remark that
$\sigma$ and $\chi$, in general, are complex fields.  However, complex
field values typically lead to large CP violation, making them
phenomenologically unacceptable; this is the reason why we restricted
our analysis to real field values.  If a non-trivial phase between the
octet and singlet condensates is favored in case of non-vanishing
$|\chi|$, this may lead to an argument against the formation of color
octet condensates in general. In order to demonstrate this point we
assume for a moment that the effective potential for the relative
phase between $\chi$ and $\sigma$ is dominated by the small field
instanton contribution $U\approx \zeta V(\sigma,\chi)$.
 Then, real positive values of $\sigma$ would be preferred
due to the instanton contribution. This would in turn lead to a
positive ``mass term'' $\sim \chi^2$ 
(cf. Fig.~\ref{largesigma}), originating from
the $\sigma \chi^2$ term in \eqref{Vpot}; for imaginary $\chi=\ii
|\chi|$, this turns into $-\sigma |\chi|^2$. Combining this with the small 
$\chi^3$ term, the relative minimum of $V(\sigma,\chi)$ for fixed 
$\sigma>0, |\chi|>0$ would occur for a complex CP-violating $\chi$.
Unfortunately, the impact of this observation is weakened by the very
small coefficients of the $\sigma\chi^2$ and $\chi^3$ terms arising in
our approximation.  The approximate flatness in the $\chi$ direction
makes the potential influence of other effects large. In this context
we observe that the $U(1)_A$-conserving part $U_0$ in \eq{effact} also
contributes to the effective potential for the phase between $\chi$
and $\sigma$, for example with terms $\sim {\sigma^*}^2 \chi^2+c.c.$.
Only the common phase of $\chi$ and $\sigma$ is protected by the
$U(1)_A$-symmetry and is uniquely determined by the instanton part.

\subsection{Beyond small condensates}

As demonstrated in the preceding section, the instanton-induced
effective potential can, in principle, support a mechanism for
spontaneous color-octet condensation. Whether or not this mechanism is
realized, however, requires a study that is valid for larger values of
$\sigma$ and $\chi$.  The consequences of large condensates are
twofold. First, the fermion masses are no longer small. This affects
the non-zero-mode contribution $\zeta_{\textrm{n}}$ as well as the
running of the gauge coupling. Also a mixing between zero modes and
non-zero modes is induced. Second, a color non-singlet field gives an
effective mass to the gauge fields, which again modifies the running
of the gauge coupling (now the pure gauge contribution). In addition,
it provides for an effective infrared cutoff for the $\rho$
integration.

\subsubsection{Effects on the running gauge coupling}

Fermion and effective gauge boson masses exert an immediate influence
on the running of the gauge coupling. For momenta smaller than the
mass of a given quark or gluon degree of freedom, the corresponding
fluctuations of this degree of freedom are suppressed. As a
consequence, these fluctuations do no longer contribute to the running
of the coupling.  This decoupling of massive modes can directly be
implemented in the $\beta$ function for the running coupling, which we
write as
\begin{equation}
\partial_{t}g^{2}\equiv k\frac{d}{dk} g^2=
-\frac{1}{8\pi^2}\,g^4\left(\frac{11}{3}\Nc\,
l_{\text{g}}(\frac{M^{2}_{g}}{k^2})-\frac{2}{3}\Nf\,
l_{\text{f}}(\frac{|M_{8}|^{2}}{k^2})\right), \quad t\equiv \ln
\frac{k}{\Lambda},
\label{gaugeflow}
\end{equation}
where $M_{g}$ and $M_{8}$ are the gluon and the octet masses given in
Eqs.~\eqref{fermionmasses},\eqref{gluonmass}, and $k$ denotes an RG momentum scale. The
threshold functions $l_{\text{g,f}}(x)$ approach unity for small
argument, $l_{\text{g,f}}(0)=1$, corresponding to the fact that the
physical or effective masses play no role in the UV $k\to \infty$. For
large argument, i.e., for momentum scales $k$ below a given mass, the
threshold functions drop to zero rapidly, $l_{\text{g,f}}(x\gg 1) \to
0$, which implements the decoupling of massive modes from the
renormalization flow. The threshold functions are not universal but
regularization scheme dependent. For generic mass-dependent schemes,
the threshold functions interpolate smoothly between the two
limits.\footnote{For mass-independent schemes such as the
  $\overline{\text{MS}}$ scheme, threshold functions do not appear
  directly; but in order to describe the physics above and below a
  mass threshold adequately, theories with the correspondingly
  different particle content have to be matched at the mass threshold.
  This can equally be described by an effective threshold function
  which changes its slope discontinuously at a mass threshold. } For
the explicit computations, we set the threshold functions equal,
$l_{\text{g}}(x)=l_{\text{f}}(x)=l(x)$, and use
\begin{equation}
\label{threshold}
l(x)=\frac{1}{(1+x)^3}.
\end{equation}
This is a typical form for a threshold function, occurring in
calculations based on the functional RG.  Of course, the one-loop form
used in Eq.~\eqref{gaugeflow} only serves as an example. A similar
analysis of mass threshold behavior applies to any loop order and even
fully non-perturbatively.  We would like to stress that it is this
threshold behavior where the additional free parameter $Z_\chi$ enters
via $M_g$, cf. Eq.~ \eqref{gluonmass}.

As a result of this decoupling mechanism, the effective running
coupling is now field dependent, $g(k,\sigma,\chi)$. Inserting this
into Eqs.~\eqref{nonzeromodes}, \eqref{potential} results in an
additional field dependence of the effective potential. Qualitatively,
the gauge boson mass weakens the increase of the gauge coupling. Owing
to the exponential of the classical action $\sim \exp(-{8\pi^2}/{g^2})$
in Eq.~ \eqref{nonzeromodes}, this leads to a total suppression of the
instanton contribution. The fermion threshold behavior has the
opposite effect due to the minus sign in the $\beta$ function,
reflecting their charge-screening nature.

\subsubsection{Effect on the instanton determinant}

\begin{figure}[t]
\begin{center}
\scalebox{0.65}[0.65]{
\begin{picture}(190,140)(40,0)
\includegraphics[width=9.5cm]{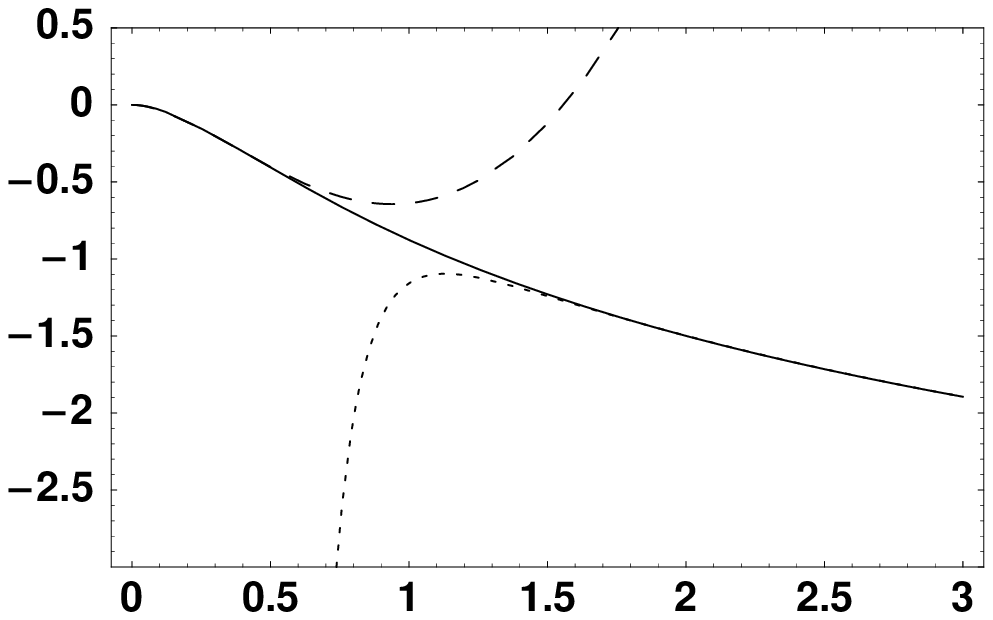}
\Text(-10,-10)[c]{\scalebox{1.6}[1.6]{$x$}}
\Text(-290,150)[c]{\scalebox{1.6}[1.6]{$K(x)$}}
\end{picture}
}
\end{center}
\vspace{-0.3cm}
\caption{The solid line gives the interpolating function $K(x)$,
  smoothly connecting the small mass (dashed) and the large mass
  approximations \cite{Dunne:2004sx,Dunne:2004cp,Kwon:2000ip}.}
\label{fitfunction}
\end{figure}

The condensates give masses to fermions and gluons, hence the
corresponding fluctuation determinants have to be evaluated for this
massive case.  Let us first consider the massive fermion determinant,
i.e., the non-vanishing shift of the fermionic (non-zero) eigenmodes
due to the effective fermion mass Eq.~\eqref{fermionmasses}. This
problem has been solved recently using an efficient method to perform
the mode sum \cite{Dunne:2004cp,Dunne:2004sx}. The result 
interpolates smoothly between the analytically known small and large
mass expansions \cite{Carlitz:1978yj,Kwon:2000ip}. These calculations
have been performed with a color singlet quark mass $m$ in the MS
scheme.  Here, we neglect the difference in the effect of the singlet
and octet quark mass and approximate 
\begin{equation}
m=\frac{1}{3}\sqrt{|M_{1}|^2+8|M_{8}|^{2}}.
\end{equation}
For our purposes, we have to adapt the results of
\cite{Dunne:2004sx,Dunne:2004cp,Kwon:2000ip} to our massive
RG regularization scheme, as derived in
Appendix~\ref{assembling}, and use the following interpolating
function
\begin{eqnarray}
\label{massbehavior}
\Nf K(x)&:=& \ln \det{}'(-\fssd{D}\,+m)\Big|_{\rm RG}\nonumber\\
&=&-\frac{2}{3}\Nf (H(x)+\frac{3}{4})+\ln\det{}'(-\fssd{D}\,+m)
\Big|_{\rm MS}
\\\nonumber
&\simeq&\Nf\left[-\ln(x)-a_{1}+\frac{\ln(x)+a_{1}-
a_{2}x^2-a_{3}x^4}
  {1+a_{4}x^2+a_{5}x^4+a_{6}x^6}\right],
\end{eqnarray}
with $x=\rho m$. The function $H(x)$ is defined in
Eq.~\eqref{massiverg}, and
\begin{eqnarray*}
a_{1}=0.792,\,\,\,\,\,a_{2}=3.58,\,\,\,\,\,
a_{3}=0.0842,\,\,\,\,\,a_{4}=0.00115,\,\,\,\,\,
a_{5}=23.5,\,\,\,\,\,a_{6}=9.28,
\end{eqnarray*}
The primed determinant $\det{}'$ in \eq{massbehavior} is that in the
space of non-zero modes. As shown in Fig.~\ref{fitfunction}, this
function interpolates smoothly between the small- and large-mass
regimes.

A similar behavior as for the fermion determinant is expected for the
non-vanishing gluon mass in the gluon determinant. However, this
effect is sub-leading, the dominant effect being the modification of
the classical action at the minimum, see e.g.\
\cite{'tHooft:fv,Pawlowski:1996ch}: for constant $\chi$, this gives a
contribution to the classical action $\Delta S_{\textrm{cl}}=-6\pi^2
Z_{\chi}|\chi|^2\rho^2$ and therefore a factor of (cf.
Eq.~\eqref{Uinst})
\begin{equation}
\label{higgs}
\exp(-8\pi^2/g^2(\rho)) \, \Delta_{\text{gl}}(\chi,\rho) \to
\exp(-8\pi^2/g^2(\rho)-6\pi^2 Z_{\chi}|\chi|^2\rho^2) \,
\Delta_{\text{gl}}(\rho) 
\end{equation}
in the integral \eqref{potential}. To summarize, the full inclusion
of $\sigma$ and $\chi$ in the fermion determinant and the Higgs-type
of contribution to the classical action result in our final formula
for the effective potential ($\Nc=\Nf=3$):
\begin{eqnarray}
U_{\textrm{inst}}\!\!&=&\!\!D_{\overline{\textrm{RG}}}V(\sigma,\chi)\int
d\rho\,\rho^{-2} \left(\frac{8\pi^2}{g^{2}(\rho)}\right)^{6}
\exp\left(-\frac{8\pi^2}{g^{2}(\rho)}-6\pi^2 Z_{\chi}|\chi|^2\rho^2+3
K(m\rho) \right), \nonumber\\ &&\label{finalresult}
\end{eqnarray}
where $K$ is given in Eq.~\eqref{massbehavior}, and
$D_{\overline{\text{RG}}}$ is defined in Eq.~\eqref{constants}. Once
the running of the gauge coupling is specified, e.g., using the
one-loop form of Eq.~\eqref{gaugeflow} and identifying the RG scale
with the inverse instanton radius, $k=1/\rho$, we can investigate the
landscape of the instanton-induced contribution to the effective
potential for $\sigma$ and $\chi$. For fixed $\sigma$, an additional
$\chi$ dependence arises from the explicit term $\sim|\chi|^2$ in the
``classical part'', the dependence of $g(\rho)$ on $\chi$ and the
threshold effect $K(m\rho)$. Our approximation of $K(m\rho)$ reflects
probably only poorly the dependence on the ratio $\chi/\sigma$, and we
have also neglected the mixing between the fermionic zero modes and
non-zero modes which would modify $V(\sigma,\chi)$.

\section{Asymptotic behavior of the effective potential}\label{asymptotic}

In order to obtain a more analytic understanding of the effective
potential $U_{\text{inst}}$, let us investigate its asymptotic
behavior for the different regimes of small and large fields $\sigma$
and $\chi$. Of particular interest is the interplay between this
asymptotic behavior and the running of the gauge coupling. As an
important {\em caveat}, it should be kept in mind that our derivation
of the effective potential is based on the semi-classical instanton
gas approximation. This approximation implicitly assumes that the
one-instanton contribution is small, which translates into a small
value of $U_{\text{inst}}$. Therefore, whenever a large asymptotic
behavior of $U_{\text{inst}}$ is encountered, this may not necessarily
reflect the true behavior but rather signal the breakdown of the
instanton-gas approximation.

Our derivation of Eq.~\eqref{finalresult} so far made use of the
specific one-loop running of the gauge coupling given in
Eq.~\eqref{gaugeflow}. Assuming that the functional dependence on the
running coupling holds also in the general case, we use the form of
Eq.~\eqref{finalresult} also for other theoretically or
phenomenologically motivated running gauge couplings. For
definiteness, we will use gauge couplings with the following infrared
($\rho\rightarrow\infty$) properties
\begin{eqnarray}
\label{asympt}
g^{2}(\rho)\Big|_{\rho\to\infty}
\sim\left\{\renewcommand{\arraystretch}{1.7}
\begin{array}{ll}
 \textrm{const}& \textrm{fixed point} \\
|\ln(\rho)|^{p_{\text{log}}} &  \textrm{logarithmic divergence}\\
\rho^{p_{\text{power}}} & \textrm{power law divergence} \\
 g^{2}_{\textrm{pert}}(\rho) \Theta(\Lambda_{\textrm{QCD}} -\frac{1}{\rho})
  +\Theta(\frac{1}{\rho}-\Lambda_{\textrm{QCD}})\infty
  & \textrm{perturbative}
\end{array}\renewcommand{\arraystretch}{1} \right\},
\end{eqnarray}
with positive constants $p_{\text{log}}$ and $p_{\text{power}}$; each
infrared behavior will be adapted to show the same decoupling
properties for massive modes as displayed in Eqs.  \eqref{gaugeflow}
and \eqref{threshold}. For simplicity, we assume that this IR behavior
does not depend on the number of fermions. These running couplings and
the corresponding resulting instanton densities at vanishing external
fields are shown in Fig.~\ref{density} in Appendix~\ref{runcoup}.

The effective potential for non-vanishing $\sigma$ and $\chi$ is
strongly influenced by fermion and gauge boson mass effects,
respectively. Hence, the effective potential is expected to behave
differently in the various directions of the $\sigma,\chi$ plane. For
definiteness, let us investigate four cases: (1)
$\sigma\,\,\textrm{small},\, \chi=0$; (2)
$\sigma\rightarrow\infty,\,\chi=0$; (3)
$\sigma=0,\,\chi\,\,\textrm{small}$ and (4)
$\sigma=0,\,\chi\rightarrow\infty$.

The $\rho$ integration can be performed analytically by splitting the
integration domain into several intervals, each of them dominated by a
different effect. For case (1), for instance, there is a UV regime
$(0,\rho_{\textrm{pert}})$ where the running coupling is well
approximated by one-loop perturbation theory without the influence of
mass thresholds. In the consecutive interval,
$(\rho_{\textrm{pert}},\frac{c}{\sigma})$ (with $c$ being some
constant $O(1)$) the gauge coupling is dominated by non-perturbative
dynamics, but fermion masses still do not play an important role.
Finally, in the interval $(\frac{c}{\sigma},\infty)$ the fermions are
heavy compared to the scale $1/\rho$, and pure gluodynamics dominates
the running coupling.  Neglecting logarithmic dependencies on $\sigma$
or $\chi$, we find,
\begin{eqnarray}
\label{asymptotics}
U_{\text{inst}}(\sigma,\chi)\sim
\left\{\begin{array}{ll}
 A_{1}\sigma^{\Nf}+B_{1}\sigma^{4+2\Nc{p_{\text{power}}}},
\qquad& (1):\,\,\sigma\,\,
 \text{small}, \,\, \chi=0
\\
A_{2}\sigma^{4-\beta_{0}}+B_{2}\sigma^{4(1-\frac{\beta_{0}}{
\beta^{\prime}_{0}})}
,\qquad& (2):\,\,\sigma\to\infty, \,\, \chi=0 \\
A_{3}\chi^{\Nf}+B_{3}\chi^{4+\frac{4\Nc{p_{\text{power}}}}{
2+{p_{\text{power}}}}}
,\qquad& (3):\,\,\sigma=0, \,\, \chi\,\,\text{small} \\
\\
A_{4}\chi^{4-\beta_{0}}
,\qquad& (4):\,\,\sigma=0, \,\, \chi\to\infty
\end{array}
\right. ,
\end{eqnarray}
where all terms with coefficients $A_i$ arise from the perturbative
interval $\rho\in(0,\rho_{\textrm{pert}})$, and those with
coefficients $B_i$ arise from the various non-perturbative intervals.
Of course, only the $B_i$ terms depend on the form of the running
coupling specified in Eq.~\eqref{asympt}; in fact, $B_{1}$ and $B_{3}$
vanish for the ''perturbative'' gauge coupling of Eq.~\eqref{asympt}.
Furthermore, $\beta_{0}=\frac{11}{3}\Nc-\frac{2}{3}\Nf$ is the
one-loop coefficient of the $\beta$ function with fermions, and
$\beta^{\prime}_{0}=\frac{11\Nc}{3}$ denotes the one-loop coefficient
for pure gluodynamics.  Not surprisingly, the logarithmically divergent and the
fixed-point coupling yield the same results on this level of accuracy;
hence the parameter $p_{\text{log}}$ of Eq.~\eqref{asympt} does not
enter Eq.~\eqref{asymptotics}.

For $\Nf> 4+2\Nc{p_{\text{power}}}$, the small-$\sigma$ and
small-$\chi$ behavior may significantly be modified compared to the
perturbative expectation $\sigma^{\Nf},\chi^{\Nf}$. Here the possible
dependence on the infrared behavior of the gauge coupling appears to
be important.  However, this simply reflects the fact that naive IR
convergence in the $\rho$ integration is lost for $\Nf>4$, as we have
already noted before. In this case, the convergence is now restored by
a combination of the suppression due to the finite fermion mass
$\exp(-\Nf K(m\rho))$ and the suppression from the gauge coupling
$\left(\frac{8\pi^2}{g^2(\rho)}\right)^{2\Nc}$ at the expense of a
direct dependence on the IR behavior of the gauge coupling.

In the large-field regime, the potential grows faster than
$\sigma^4,\chi^4$ only if $\beta_{0}<0$ which corresponds to theories
without asymptotic freedom. Conversely, for asymptotically free
theories, the instanton-induced potential will not dominate over the
non-anomalous part of the potential $U_0$ which can be expected to
exhibit a $\sigma^4,\chi^4$ growth for reasons of universality. In
view of the {\em caveat} mentioned in the beginning of this
subsection, we interpret this result as a successful self-consistency
check of the instanton-gas approximation. Furthermore, for $\Nc=\Nf=3$
and fixed $\sigma$, the instanton contribution vanishes
$U_{\text{inst}}\sim \chi^{-5}$ for large $\chi$. The instanton
potential is therefore stabilized in the $\chi$ direction.

\section{Numerical analysis of the effective potential}\label{solution}

Guided by the analytic knowledge obtained so far for the effective
potential, let us study our full result for the instanton-induced
effective potential $U_{\text{inst}}(\sigma,\chi)$ obtained
numerically from Eq.~\eqref{finalresult}. For definiteness, we use --
as an example -- a one-loop form for the gauge coupling modified such
that it approaches an IR fixed point at $g^2_{\textrm{fix}}=100$ in
absence of condensates; this is in the ball park of IR results from RG
flow equations \cite{Gies:2002af}. Our conclusions remain similar for
all other running couplings proposed in Eq.~\eqref{asympt}.
  
\begin{figure}[t]
\begin{center}
\subfigure[]{\scalebox{0.7}[0.7]{
\begin{picture}(190,140)(40,0)
\includegraphics[width=9.5cm]{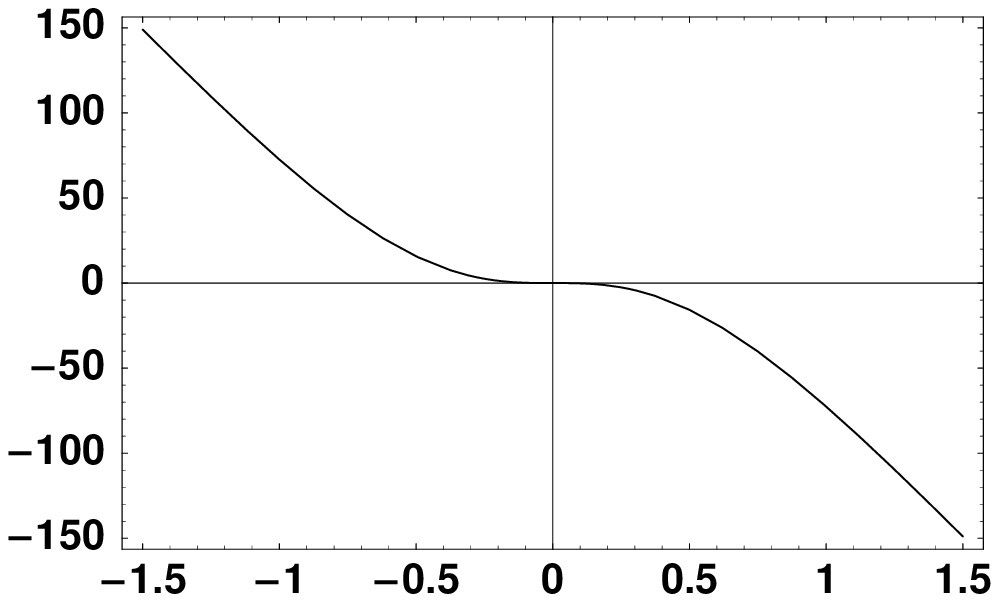}
\Text(-30,-10)[c]{\scalebox{1.6}[1.6]{$\sigma/\Lambda_{\textrm{QCD}}$}}
\Text(-290,150)[c]{\scalebox{1.6}[1.6]{$\frac{U(\sigma)}{D_{\textrm{S}}}$}}
\end{picture}
}
\label{sigma}}
\hspace{2.8cm}
\subfigure[]{\scalebox{0.7}[0.7]{
\begin{picture}(190,140)(40,0)
\includegraphics[width=9.5cm]{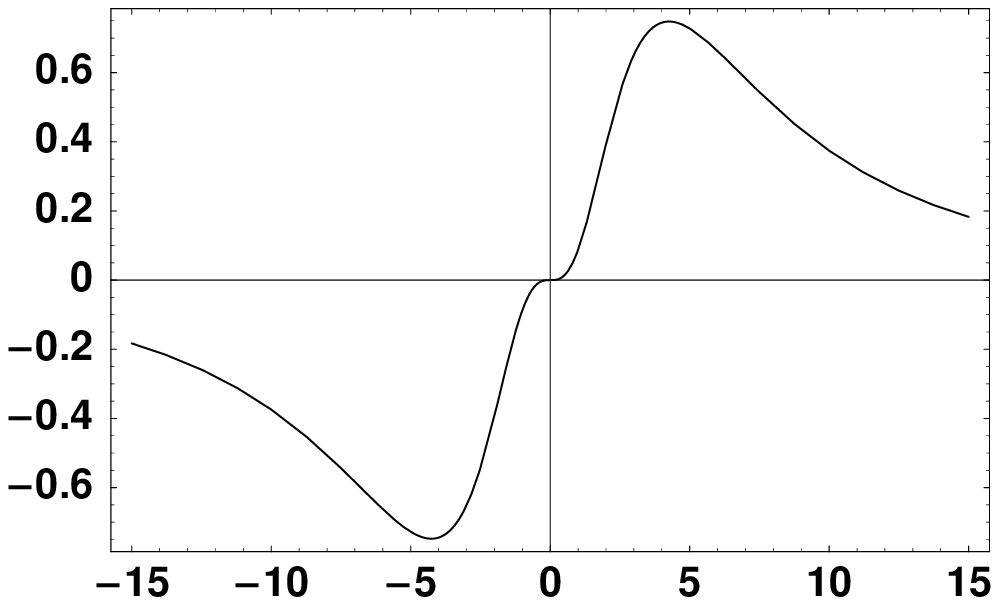}
\Text(-30,-10)[c]{\scalebox{1.6}[1.6]{$\chi/\Lambda_{\textrm{QCD}}$}} 
\Text(-290,150)[c]{\scalebox{1.6}[1.6]{$\frac{U(\chi)}{D_{\textrm{S}}}$}}
\end{picture}
}\label{chi}}
\end{center}
\vspace{-0.3cm}
\caption{Sections of the instanton potential
  $U_{\textrm{inst}}(\sigma,0)$ (left panel) and
  $U_{\textrm{inst}}(0,\chi)$ (right panel) in units of
  $\Lambda_{\textrm{QCD}}$. The different large-field behavior is
  clearly visible: the potential in the positive $\sigma$ direction
  goes to $-\infty$ whereas it rapidly approaches $0$ for large
  $|\chi|$.  Note the different scales of the potential
  itself for the $\sigma$ and the $\chi$ direction.
  (Both plots use $Z_{\chi}^{1/2}=1/15$ and a gauge coupling
  approaching an IR fixed point $g^{2}_{\textrm{fix}}=100$. The
  renormalization-scheme-dependent constant $D_{\rm{S}}$ is scaled
  out.)} 
\label{pictures}
\end{figure}

Let us first confirm the asymptotic behavior obtained analytically
above along the $\sigma$ and $\chi$ axes. As is visible in
Fig.~\ref{sigma}, displaying $U(\sigma,0)$, the potential along the
$\sigma$ axis is unbounded from below for $\sigma\rightarrow\infty$.
In particular, for $\Nf=\Nc=3$, the resulting asymptotics of
Eq.\eqref{asymptotics} yielding
$|U(\sigma,0)|\sim\sigma^{\frac{8}{11}}$ is confirmed.  Let us stress
that the overall sign of the potential is negative for positive
$\sigma$, since the integral in Eq.~\eqref{finalresult} is always
positive and the prefactor $V(\sigma,0)=-\sigma^3$ is negative. From
this, we draw two conclusions: first, the instanton potential favors
chiral symmetry breaking, but the value of the condensate is not
determined by the instanton potential alone (at least in our simple
one-instanton approximation). Second, the complete instanton potential
$U_{\mathrm{inst}}(\sigma,\chi)$ cannot have a global minimum since
there is a direction in which the potential always decreases.

Next we consider a pure $\chi$ field. Figure \ref{chi} shows that the
instanton potential becomes flat rather rapidly for large $\chi$, as
expected from Eq.~\eqref{asymptotics}. Negative $\chi$ are clearly
preferred.\footnote{Let us stress that the relative sign of $\sigma$
  and $\chi$ is indeed important, since it changes the parity of some
  particles in the spectrum of the model \eqref{model}. Moreover,
  owing to the U(1)${}_{\textrm{A}}$ anomaly, it is not clear whether
  the sign in the Yukawa couplings can be rotated away by a chiral
  transformation.} From Fig.~\ref{sigma}, we can read off the location
of the minimum: $\chi_{\textrm{min}}\sim 4.26\Lambda_{\textrm{QCD}}\sim
1.4\textrm{GeV}$ for $\Lambda_{\textrm{QCD}}=330\textrm{MeV}$. This
results in an encouraging $M_{g}\sim 350\textrm{MeV}$.

Nevertheless, it is important to observe that the potential in the
$\chi$ direction is rather shallow compared to the $\sigma$ direction,
the relative height being $\sim 10^{-3}$. This is a direct consequence
of the relative prefactors in the potential $V(\sigma,\chi)$ of Eq.~
\eqref{Vpot}. The dependence of the absolute value of $\chi$ on our
remaining free parameter $Z_{\chi}$ is strong, whereas it remains weak
for $M_{g}$. This is in agreement with the expectation that, in the
octet direction, the threshold effect of the gluon mass $M_{g}$ is
much stronger than that of the quark masses $\sim \chi$. For our
quantitative results, we use the value $Z_\chi^{1/2}=1/15$ which is in
the phenomenologically acceptable range \cite{Wetterich:2000pp,35A}.

Finally, let us study the complete potential depending on both fields
$\sigma$ and $\chi$.  Even though we have already observed that the
instanton contribution alone does not have a global minimum, it is
nevertheless worthwhile to look for a local one. Such a local minimum
indeed exists at $(\sigma,\chi)\approx(-0.27,-4.2)\Lambda_{\text{QCD}}$
with the absolute depth of the potential being
$U(\sigma,\chi)|_{\text{loc.  min.}}\approx -1.42 \,D_{\rm{S}}\approx
0.009$ in units of $\Lambda_{\textrm{QCD}}$ (and using
$D_{\overline{\mathrm{MS}}}\approx 6\times 10^{-3}$).  Since this
minimum has a too small $|\sigma|\sim 90\,\textrm{MeV}$ and is
extremely shallow, it is not physically acceptable. Any generic
non-anomalous contribution $U_0$ is likely to remove this minimum.

\begin{figure}[t]
\begin{center}
\subfigure[]{\scalebox{0.7}[0.7]{
\begin{picture}(190,140)(40,0)
\includegraphics[width=9.5cm]{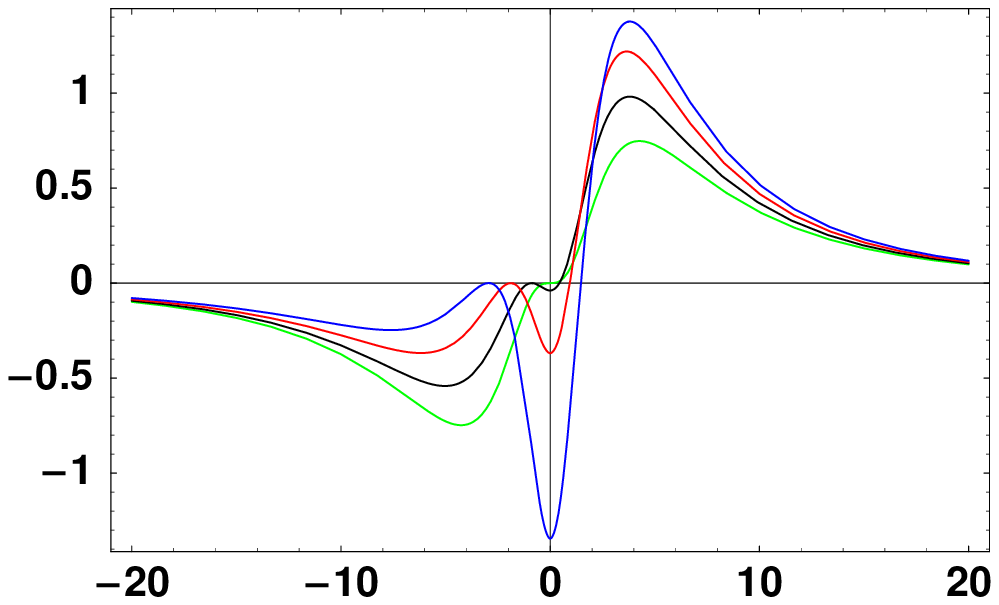}
\Text(-30,-10)[c]{\scalebox{1.6}[1.6]{$\chi/\Lambda_{\text{QCD}}$}}
\Text(-280,150)[c]{\scalebox{1.6}[1.6]{$\frac{U(\chi)}{D_{\rm S}}$}}
\end{picture}
}
\label{smallsigma}}
\hspace{2.8cm}
\subfigure[]{\scalebox{0.7}[0.7]{
\begin{picture}(190,140)(40,0)
\includegraphics[width=9.5cm]{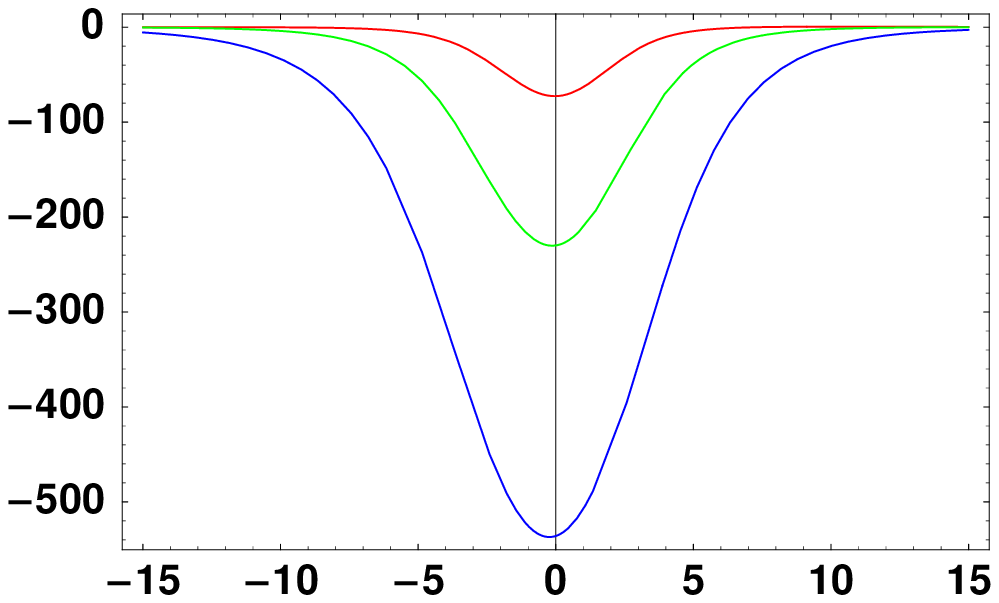}
\Text(-30,-10)[c]{\scalebox{1.6}[1.6]{$\chi/\Lambda_{\text{QCD}}$}}
\Text(-280,150)[c]{\scalebox{1.6}[1.6]{$\frac{U(\chi)}{D_{\rm S}}$}}
\end{picture}
}\label{largesigma}}
\end{center}
\vspace{-0.3cm}
\caption{Effective potential $U_{\text{inst}}(\sigma,\chi)$ for different
  fixed $\sigma$. The left panel displays the cases of $\sigma=0,\,
  0.06,\, 0.128,\, 0.20\Lambda_{\text{QCD}}$ in green, black, red and 
  blue, respectively.  We observe that for values of
  $\sigma>\sigma_{\textrm{crit}}\approx 0.128\Lambda_{\text{QCD}}$
  (red curve) the trivial minimum in the $\chi$ direction becomes the
  global minimum. The right panel shows the potential for realistic
  values of $\sigma=1,2,4\Lambda_{\text{QCD}}$ in red, green and blue,
  respectively (descending order), clearly demonstrating the absence
  of an instanton-induced color octet condensate. The parameters are
  chosen as in Fig.~\ref{pictures} and all numbers refer to units of
  $\Lambda_{\text{QCD}}$.}
\label{pictures2}
\end{figure}

Since the local minimum is not acceptable and a purely
instanton-induced global minimum does not exist, let us redo our
analysis with one additional assertion: we assume that the
non-anomalous contribution $U_0(\sigma,\chi)$ to the effective
potential supports scalar singlet condensation via spontaneous
symmetry breaking for sufficiently strong gauge coupling (in fact,
this has been shown to happen generically in QCD-like theories in
\cite{Gies:2002hq}).  In order to introduce as few parameters as
possible, we simply assume that $U_0$ fixes a non-zero value
of $\sigma$, leaving the detailed form of $U_0$ aside. Now, since
$U_0$ is of non-anomalous origin, its form reflects the full
chiral symmetry (even though it can exhibit a symmetry-breaking
minimum), implying its invariance under $\sigma\to -\sigma$. Hence,
$U_0$ does not prefer a particular sign of $\sigma$.  By contrast,
$U_{\text{inst}}$ does prefer positive values of $\sigma$ as displayed
in Fig.~\ref{sigma}.

The instanton potential for small positive $\sigma$ is depicted in
Fig.~\ref{smallsigma}. We observe that the global minimum for $\chi$
is non-vanishing only for very small values of $\sigma$.  Beyond the
critical value of $\sigma_{\textrm{crit}}\approx
0.128\Lambda_{\text{QCD}}<50\textrm{MeV}$ the global minimum is at
$\chi=0$. This holds, in particular, for realistic values of
$\sigma=1\dots 4\Lambda_{\text{QCD}}$, which is plotted in
Fig.~\ref{largesigma}. For $\sigma \gtrsim \Lambda_{\rm QCD}$ the
dominant feature of the $\chi$ dependence of $U_{\rm inst}$ is simply
the vanishing of $U_{\rm inst}$ for large $\chi$ which results in a
relative minimum at $\chi=0$. Figure \ref{largesigma} summarizes one
of our main results, namely, that the instanton-induced potential
appears incapable of giving rise to an octet condensate in the present
instanton-gas approximation.

\begin{figure}[t]
\begin{center}
\subfigure[]{\scalebox{0.7}[0.7]{
\begin{picture}(190,140)(40,0)
\includegraphics[width=9.5cm]{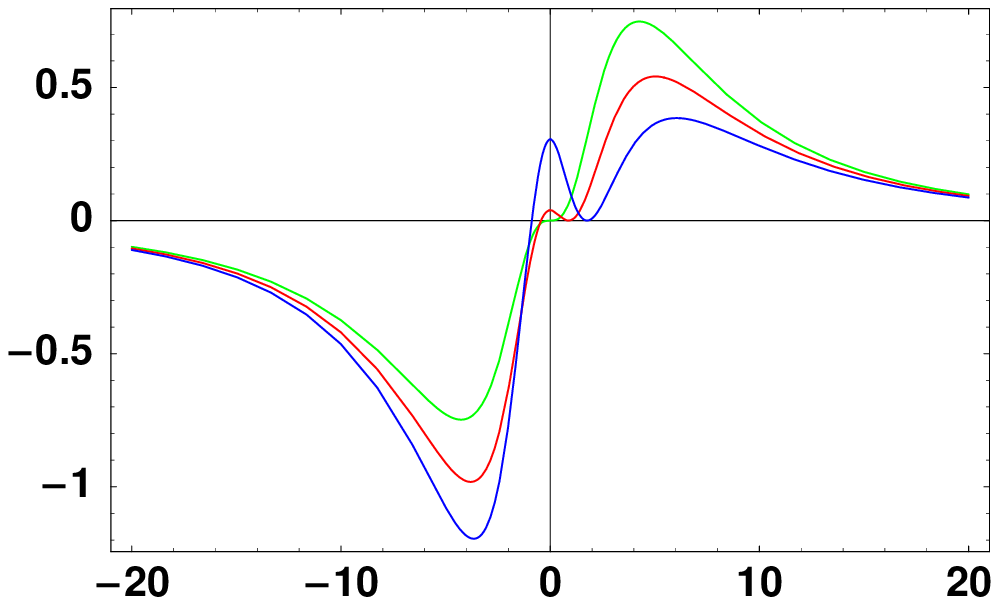}
\Text(-30,-10)[c]{\scalebox{1.6}[1.6]{$\chi/\Lambda_{\text{QCD}}$}}
\Text(-290,150)[c]{\scalebox{1.6}[1.6]{$\frac{U(\chi)}{D_{\rm S}}$}}
\end{picture}
}\label{smallsigmanegative}}
\hspace{2.8cm}
\subfigure[]{\scalebox{0.7}[0.7]{
\begin{picture}(190,140)(40,0)
\includegraphics[width=9.5cm]{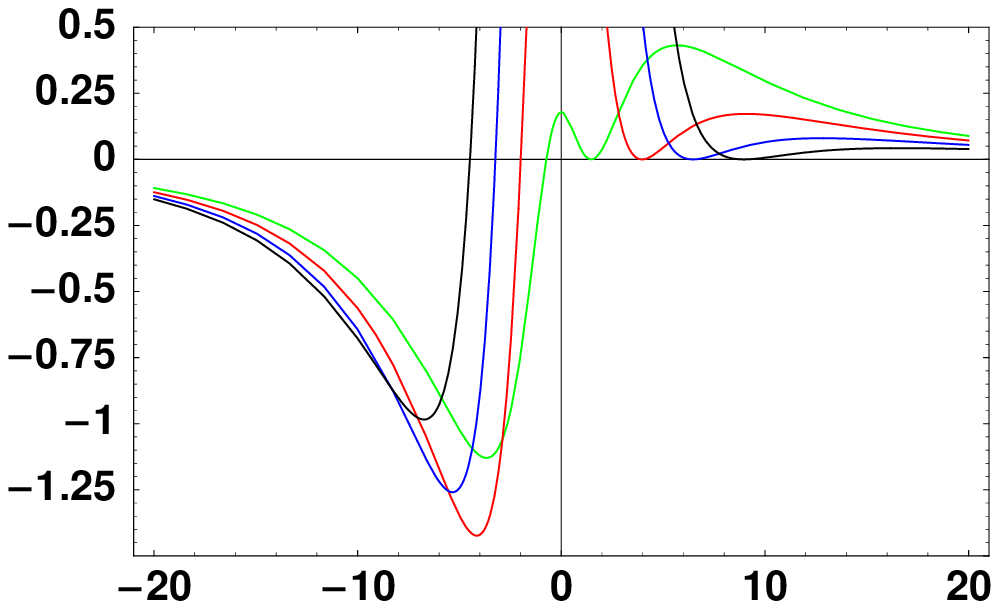}
\Text(-30,-10)[c]{\scalebox{1.6}[1.6]{$\chi/\Lambda_{\text{QCD}}$}}
\Text(-290,150)[c]{\scalebox{1.6}[1.6]{$\frac{U(\chi)}{D_{\rm S}}$}}
\end{picture}
}\label{sigmanegative}}
\end{center}
\vspace{-0.3cm} \caption{ $U_{\text{inst}}(\sigma,\chi)$ for negative
  fixed $\sigma$. On the left panel, we choose
  $\sigma=0,\,-0.06,\,-0.12\Lambda_{\text{QCD}}$ (green, red and
  blue). The global minimum then is always at non-vanishing (negative)
  $\chi$. On the right panel, we depict the behavior for somewhat more
  negative $\sigma=-0.1,-0.27,-0.44,-0.61\Lambda_{\text{QCD}}$
  (green, red, blue and black).  After increasing up to
  $\sigma\approx-0.27\Lambda_{\text{QCD}}$ (red curve) the minimum in
  the $\chi$ direction becomes more and more shallow with decreasing
  $\sigma$. The parameters are chosen as in Fig.~\ref{pictures}.}
\label{pictures3}
\end{figure}

Incidentally, the us study the potential also for negative fixed
$\sigma$. Even though positive values are clearly preferred by
$U_{\text{inst}}$, observation only constrains the modulus of $\sigma$
to be in the realistic range $|\sigma|\sim 1\dots 4$. Negative sigma
indeed always give rise to a global minimum at $\chi\neq 0$, i.e.
supporting a color octet; see Fig \ref{pictures3}. However, for
realistic values of $\sigma$, the instanton-induced potential is
extremely shallow again, such that this minimum is likely to be washed
out by the non-anomalous part (unless the latter is either fine-tuned
or supports an octet condensate itself).

\section{Conclusions}
\label{sec:conc}

We have calculated the one-instanton contribution to the effective
potential in a background of classical bosonic fields coupled to
quarks and gluons.  One field, $\sigma$, has the structure of the
typical singlet chiral condensate and the other, $\chi$, exhibits a
color-flavor-locking structure, as conjectured for the scenario of
spontaneous color symmetry breaking in the vacuum.  Beyond leading
order in the background fields, we have included effects of quark
masses on the running gauge coupling and the quark determinant.  In
addition, the color octet condensate works as a Higgs field for the
gluons, providing for an extra $\chi$-dependent contribution to the
classical instanton action.  We work in a massive regulator scheme
which makes threshold behavior more transparent.

For the realistic case of $\Nf=3$ light quark flavors, the instanton
potential is unbounded from below for a pure singlet chiral
condensate, favoring a non-trivial value of this condensate and chiral
symmetry breaking. Although there exists a local extremum with
non-vanishing octet contribution, it is rather shallow and thus likely
to be washed out by non-instanton effects. Moreover, this extremum has
a nearly vanishing value for the chiral condensate, making it
phenomenologically unacceptable. As the potential is unbounded from
below along the $\sigma$ direction, a global minimum of the instanton
potential alone is excluded. Other stabilizing effects are typically
expected from the U(1)${}_{\textrm{A}}$-preserving sector. Therefore,
we have investigated if a color octet condensate is favored at fixed
singlet chiral condensate. For realistic positive values of the chiral
field $\sigma\sim 1\Lambda_{\textrm{QCD}}$, $\chi=0$ is the global
minimum in the color octet direction. For negative $\sigma$ with a
similar absolute value, there exists a minimum with non-vanishing
octet condensate.  However, this minimum is also unnaturally shallow
and thus will presumably be washed out by non-instanton effects.
Moreover, the negative sign of $\sigma$ is disfavored by the instanton
contribution.

An analytic insight into the effective potential can be gained from
its asymptotic behavior for large background fields. It is interesting
to note that the behavior for very large fields is inherently
connected to asymptotic freedom. For instance, large color octet
condensates lead to a high-scale decoupling, such that any
non-perturbative IR behavior is screened; in this large-$\chi$
direction, the flattening of the potential can directly be related to
the perturbative approach to asymptotic freedom. Another extreme
example is provided by the case where asymptotic freedom is lost,
e.g., owing to too many fermion species; in this case, the potential
may go to $-\infty$ faster than the fourth power of the fields,
preventing stabilization by a renormalizable non-instanton potential
for the bosonic fields. On the other hand, we observe a strong
qualitative dependence on the non-perturbative IR behavior of the
coupling only for very small fields and more than four light fermion
species: here, the increase of the instanton amplitude depends
strongly on the infrared details of the gauge coupling and may be
changed from the naive $\sigma^{\Nf},\chi^{\Nf}$.

In conclusion, barring large higher-order, e.g., multi-instanton,
effects, we find that chiral symmetry breaking is supported by
instanton effects. On the other hand, the issue of color octet
condensation remains inconclusive. In our approximation, we find no
evidence that instantons favor a color condensation in the vacuum.

\section*{Acknowledgment}
The authors are grateful to G.V.~Dunne, H.~Min, and A.~Ringwald for
discussions.  HG, JJ, and JMP acknowledge support by the DFG under Gi
328/1-3 (Emmy-Noether program).

\begin{appendix}

\section{The one-instanton approximation and the instanton gas} \label{instgas}

Here we recall the computation of the instanton-induced effective
action $\Gamma$ within the dilute-gas approximation. In a dilute and
weakly interacting instanton gas, the dominating contribution to the
generating functional is the one-instanton and one-anti-instanton
contribution. It reads
\begin{equation}
Z_{1}=-\Omega\, \left(U_{\textrm{I}}(\sigma,\chi)+ 
U^*_{\textrm{I}}(\sigma,\chi)\right)
\end{equation}
where $\Omega$ denotes the 4-Volume, and $U_{\text{I}}$ is the
effective potential corresponding to the product of functional
determinants of the fluctuating fields in this background. The
anti-instanton contributes with
$U_{\textrm{AI}}(\sigma,\chi)=U^{\star}_{\textrm{I}}(\sigma,\chi)$.
Within the dilute-gas approximation, the contribution of the
$|n|$-instanton sector is given by
$Z_{n}=\frac{(Z_{1})^{n}}{n!}$. This leads to the the full amplitude
\begin{equation}\label{ZI}
Z=\sum^{\infty}_{n=0}Z_{n}=\sum^{\infty}_{n=0}
\frac{(Z_{1})^{n}}{n!}=\exp(Z_{1}),
\end{equation}
where we have normalized the zero-instanton amplitude to one. To
lowest order in the bosonic background fields, this holds because we
included the influence of the fields (masses for the fermions) only
for the zero modes.  However, in the absence of gauge fields,
the Dirac operator has no zero modes. Beyond this approximation, this
normalization corresponds to a modification of the non-instanton
U(1)${}_{A}$-symmetric contribution to $U(\sigma,\chi)$. From \eq{ZI},
we read off the effective action in the classical
background of $\sigma$, $\chi$,
\begin{equation}
\Gamma=-\ln Z=-Z_{1},
\end{equation}
which serves as the starting point of our investigation in the main
text.

\section{Different regularization schemes}\label{rgscheme}

In this appendix, we discuss how to switch between regularization
schemes. We shall use a scheme which manifestly exhibits the
decoupling of massive modes. It has been shown in
\cite{Pawlowski:1996ch} that topological effects persist within the RG
framework used in the present work. This applies, in particular, to
the existence of zero modes \cite{Pawlowski:1996ch}. Moreover, at
leading order one only has to take into account the explicit mass or
regulator dependence. This amounts to using the well-known zero modes
\cite{'tHooft:fv}.

For non-perturbative problems involving mass
threshold effects, as they are induced by the background fields in our
case, such schemes are highly advantageous. Following
\cite{'tHooft:fv,'tHooft:1973us}, a change of the renormalization
scheme can be understood by comparing two integrals,
\begin{eqnarray}
\label{idef}
I_{1}=\int \frac{d^4q}{(2\pi)^4}\frac{1}{(q^2+m^2)^2},\quad
I_{2}=\int\frac{d^4q}{(2\pi)^4}\frac{q^4}{(q^2+m^2)^4},
\end{eqnarray}
in the different regularization schemes. Here, $I_{1}$ appears in
connection with the zero modes and $I_{2}$ with non-zero modes.
Therefore, we can easily keep track of these terms. For example,
the Pauli Villars scheme gives
\begin{eqnarray}
\label{pauli}
I^{\textrm{PV}}_{1}&=&\int
\frac{d^4q}{(2\pi)^4}\left[\frac{1}{(q^2+m^2)^2}-\frac{1}{(q^2+
\Lambda_{\text{PV}}^2)^2}\right]
=\frac{1}{(4\pi)^2}(2\ln(\Lambda_{\text{PV}})-2\ln(m)),
\\\nonumber
I^{\textrm{PV}}_{2}&=&\int
\frac{d^4q}{(2\pi)^4}\left[\frac{q^4}{(q^2+m^2)^4}-\frac{q^4}{(q^2+
\Lambda_{\text{PV}}^2)^4}\right]
=\frac{1}{(4\pi)^2}(2\ln(\Lambda_{\text{PV}})-2\ln(m))=I^{\textrm{PV}}_{1}, 
\end{eqnarray}
whereas we find in
dimensional regularization
\begin{eqnarray}\nonumber
I^{\textrm{dreg}}_{1}&=& \mu^{4-n}\int\frac{d^nq}{(2\pi)^n}
\frac{1}{(q^2+m^2)^2}\nonumber\\
&=&\frac{1}{(4\pi)^2}
  \left[\frac{2}{4-n}+2\ln(\mu)-2\ln(m)-\gamma+\ln(4\pi)+O(4-n)\right]\\
&=&I^{\textrm{dreg}}_{2}+\frac{5}{96\pi^2}.
\label{dimreg}\end{eqnarray}
Comparing Eqs.~\eqref{pauli} and \eqref{dimreg}, the substitutions for
a change from Pauli-Villars to dimensional regularization read
\begin{eqnarray}
\label{transition_1}
\!\!\!\!\!\!\!\!&&I_{1}:\,\,\ln(\Lambda_{\text{PV}})\rightarrow
   \frac{1}{4-n}+\ln(\mu)-\frac{1}{2}\gamma+\frac{1}{2}\ln(4\pi),
\\\nonumber
\!\!\!\!\!\!\!\!&&I_{2}:\,\,\ln(\Lambda_{\text{PV}})\rightarrow\frac{1}{4-n}
+\ln(\mu)
   -\frac{1}{2}\gamma+\frac{1}{2}\ln(4\pi)-\frac{5}{12}.
\end{eqnarray}
Using Eq.~\eqref{transition_1}, it is easy to check that, starting
from Eqs.  \eqref{nonzeromodes}, \eqref{zeromodes}, we can obtain the
corresponding result in dimensional regularization as given, e.g., in
\cite{'tHooft:fv}.

Both schemes discussed so far are mass independent. This originates
from the fact that for fixed cutoff $\mu$ or fixed dimensionality
$4-n$, the integrals $I^{\textrm{PV}}$ and $I^{\textrm{dreg}}$ do not
vanish in the limit $m\rightarrow \infty$. A mass-dependent
regularization scheme should implement this decoupling: massive modes
should not contribute to physics below the mass threshold. For
$m\gg\mu$ the integrals $I_{1}$ and $I_{2}$ should become small and
vanish in the above limit $\mu$ fixed, $m\rightarrow\infty$. This can
be implemented by defining
$I^{\textrm{RG}}_{1}\stackrel{!}{=}I^{\textrm{RG}}_{2}$ with
\begin{eqnarray}
\label{rgreg}
I^{\textrm{RG}}_{1}
&=&\int^{\Lambda}_{0}dk\,k^{-1}\, l\left(\frac{m^2}{k^2}\right)
=\frac{1}{(4\pi)^2}\left[\ln\left(1+\frac{\Lambda^2}{m^2}\right)
  -\frac{2m^2\Lambda^2+3\Lambda^4}{2(m^2+\Lambda^2)^2}\right],
\\\nonumber
l(\omega)&=&\frac{1}{(1+\omega)^3}.
\end{eqnarray}
In fact, this is not an arbitrary definition, but receives motivation
from various sources. First, it is very convenient to have
$I_{1}=I_{2}$, since, when changing from the common Pauli--Villars
scheme to our scheme it is not necessary to distinguish between the
different contributions from $I_{1}$ and $I_{2}$. The main reason,
however, is the simple form of the one-loop flow equation for
the gauge coupling in Eq.~\eqref{gaugeflow} which results from
this choice.  Finally, a deeper reason for the choice is that it
corresponds to a simplified version of a typical functional RG scheme
regularization (for more details see,
e.g. \cite{Jaeckel:2003uz}). Indeed,
\begin{equation}
{I^{\textrm{RG}}_{1}
=-\int^{\Lambda}_{0}\frac{dk}{ k}
  \int\frac{d^4q}{(2\pi)^4}\,\frac{d}{dk}R_k(q^2)
\frac{1}{(q^2+m^2+R_k(q^2))^3}, }
\end{equation}
is a typical expression for $I_{1}$ when one defines perturbation
theory from a flow equation with regulator function $R_k(q^2)$.
 In a consequent RG calculation several different
threshold functions similar to $l(\omega)$ appear. For
simplicity we put $I_{1}=I_{2}$. This is not an approximation but
simply an implicit definition of the related regulator function
$R_k$. For computations beyond the present qualitative
setting we suggest using an optimized regulator
\cite{Litim:2000ci,Litim:2001up}
\begin{eqnarray}\label{rgcutoff}
R_k(q^2)=(k^2-q^2)\,\Theta(1-\s0{q^2}{k^2})\,,
\end{eqnarray}
and its upgrades suitable for momentum-dependent approximations
\cite{Pawlowski:2005xe}.

From Eq.~\eqref{rgreg}, it is easy to find the relation between
Pauli--Villars regularization and our scheme,
\begin{equation}
\label{transition}
\ln(\Lambda_{\text{PV}})\rightarrow\ln(m)+\int^{\Lambda}_{0}\frac{dk}{k}\,
l\left(\frac{m^2}{k^2}\right)
=\frac{1}{2}\ln(m^2+\Lambda^2)
  -\frac{2m^2\Lambda^2+3\Lambda^4}{4(m^2+\Lambda^2)^2}.
\end{equation}
We emphasize that Eq.~\eqref{transition} depends on the cutoff
$\Lambda$ as well as on the mass $m$ of the particle in question. For
large $m$, the mass acts similar to the cutoff $\Lambda$. This
implements the decoupling of heavy modes.

Finally, we exploit the freedom of redefining the coupling constant at
one-loop order, such that it absorbs part of the {finite
changes discussed above},
\begin{equation}
\label{barring}
\frac{8\pi^2}{g^{2}_{\overline{\textrm{S}}}(\rho)}
=\frac{8\pi^2}{g^{2}_{\textrm{S}}(\rho)}+C_{\textrm{S}\overline{\textrm{S}}}.
\end{equation}
modifying the perturbative expression for $g^2$ only at order $g^4$.
This is often used to simplify expressions, e.g., in the transition
from $\textrm{MS}$ to $\overline{\textrm{MS}}$, or to ensure direct
comparability between different schemes. We will use this freedom
below in Appendix~\ref{assembling} to facilitate comparisons between
our scheme and the $\overline{\textrm{MS}}$ scheme in which most
results are given in the literature.

\section{Assembling the instanton integral}\label{assembling}

In this appendix, we put together all the various pieces of the
instanton size $\rho$ integral, taking care of our mass-dependent
regularization scheme. At fixed instanton size and using
Pauli-Villars regularization, the following functions contribute to
the renormalized integrand:
\begin{eqnarray}
\label{instcontrib}
f_{\textrm{cl}}&=&\exp(-S_{\textrm{cl}})=\exp(-\frac{8\pi^2}{g^{2}(\mu)}),
\\
f_{\textrm{nonzero}}&=&\exp\left(-\frac{1}{3}\Nc\ln(\mu\rho)-\alpha(1)
+\frac{\Nf}{3}\ln(\mu\rho)+2\Nf\alpha(\frac{1}{2})\right),
\\
f_{\textrm{gauge}}&=&
  \frac{4}{\rho^5}\left(\frac{4\pi}{g^2(\mu)}\right)^{2\Nc}(\mu\rho)^{4\Nc},
\\
f_{\textrm{fermion}}&=&\frac{M^{\Nf}}{\mu^{\Nf}},
\\
N&=&\frac{4}{\pi^2}\frac{\pi^{2(\Nc-1)}}{(\Nc-1)!(\Nc-2)!},
\end{eqnarray}
where $M^{\Nf}$ represents $V(\sigma,\chi)$ as defined in Eq.~
\eqref{zeromodes} and reduces to $m^{\Nf}$ if the bosonic sources only
lead to a simple mass term $m$.

Here $f_{\textrm{cl}}$ is the contribution from the classical action,
$f_{\textrm{nonzero}}$ summarizes all effects from the non-zero modes,
$f_{\textrm{gauge}}$ is the contribution from the gauge and
$f_{\textrm{fermion}}$ from the fermion zero modes. $N$ collects some
normalization factors and the group averaging. Combining all these
contributions, we find the well known result,
\begin{eqnarray}
\label{prelim_a}
N\prod_{x} f_x&=&D_{\textrm{PV}}\rho^{-5+\Nf}
\left(\frac{8\pi^2}{g^2(\mu)}\right)
  \exp\left(-\frac{8\pi^2}{g^2(\mu)}+\beta_{0}\ln(\mu\rho)\right)
\\\nonumber
&=&D_{\textrm{PV}}\rho^{-5+\Nf}
\left(\frac{8\pi^2}{g^2(\mu)}\right)
  \exp\left(-\frac{8\pi^2}{g^{2}_{\textrm{PV}}(\rho)}\right),
\end{eqnarray}
where we have used the one-loop relation between $g^2(\mu)$ and
$g^2(\rho)$ in the last step, and $D$ is given in the Pauli-Villars
scheme as,
\begin{eqnarray}
D_{\textrm{PV}}&=&
\exp\left(-\alpha(1)-2(\Nc-2)\alpha(\frac{1}{2})
          +2\Nf\alpha(\frac{1}{2})\right)=1.1506,
\\\nonumber
\alpha(\frac{1}{2})\!\!\!&=&2R-\frac{1}{6}\ln(2)-\frac{17}{72}
=0.1459, \quad \alpha(1)=8R=\frac{1}{3}\ln(2)-\frac{16}{9}=0.4433,
\\\nonumber
R\!&=&\frac{1}{12}(\ln(2\pi)+\gamma)
  +\frac{1}{2\pi^2}\sum^{\infty}_{s=2}\frac{\ln(s)}{s^2}=0.2488.
\end{eqnarray}
We point out that we still have $g^2(\mu)$ in the prefactor of the
exponential which is an artifact of the one loop calculation. This
will be rectified by higher-loop orders where the 'bare' $g^2(\mu)$ in
the prefactor is replaced by its running counterpart evaluated at the
scale $\rho$ (also at one loop order less than the corresponding one
in the exponential). Since replacing $g^2(\mu)\rightarrow g^2(\rho)$
is the main effect of higher loop orders (apart from possible changes
in the factor $D$), we account for these prefactors as well as for the
term in the exponential by hand without further calculation. In this
way, we already arrive at Eqs. \eqref{nonzeromodes}, \eqref{potential}
(at least naively in the Pauli-Villars scheme),
\begin{eqnarray}
\label{prelim}
N\prod_{x}&=&D_{\textrm{PV}}\rho^{-5+\Nf}
\left(\frac{8\pi^2}{g^{2}_{\textrm{PV}}(\rho)}\right)
\exp\left(-\frac{8\pi^2}{g^{2}_{\textrm{PV}}(\rho)}\right).
\end{eqnarray}
The final task now is to change from the mass-independent
Pauli-Villars scheme to our mass-dependent RG scheme. Starting from
Eq.~\eqref{prelim}, this is immediately done using \eqref{transition},
resulting in an additional multiplicative factor,
\begin{equation}
\label{massiverg}
RG(M_{\text{g}}\rho,m_{\text{f}}\rho)=\exp\left(\frac{11}{3}\Nc
  H(M_{\text{g}}\rho)-\frac{2}{3} 
\Nf H(m_{\text{f}}\rho)\right), \quad
H(x)=\frac{1}{2}\ln(1+x^2)-\frac{3+2x^2}{4(1+x^2)^2},
\end{equation}
where $M_{\text{g}}$ and $m_{\text{f}}$ are the gauge-boson and
fermion masses, respectively.

Finally, we make a last change and define a modified
$\overline{\textrm{RG}}$ scheme via Eq.~\eqref{barring} with
\begin{equation}
C_{\textrm{RG}\overline{\textrm{RG}}}
 =-\ln\left(\frac{RG(0,0)D_{\textrm{PV}}}{D_{
\overline{\textrm{MS}}}}\right),
\end{equation}
where $D_{\overline{\textrm{MS}}}$ is defined in \eqref{constants}.
By construction, this establishes that our coupling constant is equal
to the one-loop $\overline{\textrm{MS}}$ coupling, also including the
constant in the instanton integral. But most importantly, we have not
absorbed the mass-dependent contributions. Therefore, our scheme is
still mass-dependent and provides for decoupling of heavy modes.

\section{Calculation of the Zero-Mode Part}\label{zerodet}
The Dirac operator $\fssd{D}$\ \,in the background of an instanton has
$\Nf =3$ zero modes, being flavor copies of a fundamental zero mode. We show
that in leading order
\begin{eqnarray}\label{eq:instpot}
\zeta_{\textrm{z}}(\rho,\sigma,\chi)=
\langle \det{}_{\textrm{flavor}}
\langle\psi_0(a,i)|M_{\psi,ij}|\psi_0(b,j)\rangle\rangle_{\textrm{SU}(3)}=:
-\rho^{\Nf}V(\sigma,\chi),
\end{eqnarray}
where
\begin{eqnarray}\label{eq:M}
(M_{\psi,ij})_{\alpha\beta}=((-\fssd{D}\,_{ij})_{\alpha\beta}+
(\sigma\delta_{ij}+\chi_{ij})\delta_{\alpha\beta})\,,
\end{eqnarray}
with mass matrix $\sigma+\chi$, where $\chi$ introduces color-flavor
mixing.  The eigenvalues of $M_\psi$ are $\lambda_n(\sigma,\chi)$ with
eigenfunctions $\psi_n(\sigma,\chi)$. The color-flavor mixing
term $\chi$ does not commute with
$\fssd{D}\,$ and the $\psi_n$ are not eigenfunctions of $\fssd{D}\,$ for
$\chi\neq 0$, leaving \eq{eq:instpot} a non-trivial identity.
The mass matrix reads more explicitly
\begin{eqnarray}
(\sigma_{ab}\delta_{ij}+\chi_{ab,ij})\delta_{\alpha\beta}
=\left[\sigma\delta_{ab}\delta_{ij}
+\frac{1}{\sqrt{6}}\chi\left(\delta_{ai}\delta_{bj}-
\frac{1}{3}\delta_{ab}\delta_{ij}\right)\right]\delta_{\alpha\beta},
\label{eq:explicit}\end{eqnarray}
where we have already absorbed the Yukawa couplings into the fields
and used the color-flavor structure \eqref{condensates} for the
condensates. For clarity we have explicitly written out the spin
indices $\alpha,\beta$. Using the group averages in \eq{eq:instpot}
\cite{Shifman:uw,Creutz:mg}, we arrive at 
\begin{eqnarray}\label{eq:zeta}
\zeta_{\textrm{z}}(\rho,\sigma,\chi)=\rho^{\Nf }
\left(\sigma+\frac{1}{6\sqrt{6}} \chi\right)^2\left
(\sigma-\frac{1}{3\sqrt{6}} \chi\right).
\end{eqnarray}
It is left to prove \eq{eq:instpot}. With trivial flavor
structure, $\chi_{ij}=0$, the determinant factorizes trivially,
\begin{eqnarray}\label{eq:factorise}
\det M_\psi= \zeta_{\textrm{z}}(\rho,\sigma,\chi)\, \det{}' M_\psi\,,
\end{eqnarray}
where $\det{}'$ stands for the determinant on the non-zero mode space.
For non-vanishing $\chi$ \eq{eq:factorise} holds up to terms
$\chi^{\Nf }$. This is shown in an expansion about the determinant of
$-\fssd{D}+\sigma$ with eigenvalues
$\lambda_n(\sigma,0)=\lambda_n(0,0)+\sigma$ and eigenfunctions
$\psi(\sigma,0) =\psi_n(0,0)$. We label the zero modes of $\fssd{D}$
with $\psi_{n_0}$, $n_0=1,2,3$ with eigenvalues
$\lambda_{n_0}(s=0)=\sigma$.
There is no term linear in $\chi$ as the only invariant is $\tr_{\rm
  flavor} \chi=0$.  The quadratic term has the structure $\sigma
u_2(\sigma)\chi^2$ with finite limit $u(0)$.  It is evaluated as
\begin{eqnarray}\label{eq:chi^2}
\frac{1}{2}\left.\partial_s^2\det (-\fssd{D}+\sigma+s\chi)\right|_{s=0}
=\frac{1}{2}\sum_n (\partial^2_s \lambda_n)\,\prod_{m\neq n}\lambda_m
+\sum_{m<n}(\partial_s\lambda_n)(\partial_s\lambda_m)\,
\prod_{l\neq n,m}\lambda_l\,,
\end{eqnarray}
where the limit $s=0$ on the right-hand side of \eq{eq:chi^2} is understood.
The term proportional to $\partial^2_s\lambda_n$ has the coefficient
$\prod_{m\neq n}\lambda_m$ containing at least two of the
eigenvalues $\lambda_{n_0}(s=0)=\sigma$ of the zero modes $\psi_{n_0}$.
Hence it only contributes to sub-leading terms like $\rho \sigma^2\chi^2$.
The term proportional to $(\partial_s\lambda)^2$ has, apart from sub-leading
terms, one contribution proportional to $\prod_{l\neq n_0,m_0}\lambda_l$
removing two zero modes from the product. Thus we have
\begin{eqnarray}\label{eq:leadingfunc}
u_2(\sigma)\, \sigma^{\Nf -2}\chi^2=
\sigma^{\Nf -2} \sum_{m_0<n_0}(\partial_s\lambda_{n_0})
(\partial_s\lambda_{m_0})\,\det{}' M_\psi +O(\sigma^2\chi^2)\,,
\end{eqnarray}
where $\det{}' M_\psi=\prod_{l}{}'\lambda_l$, the primed product involves
only the non-zero eigenvalues of $\fssd{D}\,$. The $s$ derivatives in
\eq{eq:leadingfunc} follow as
\begin{eqnarray}\label{eq:id1}
\partial_s\lambda_n(s=0)=
\partial_s \langle \psi_n| (-\fssd{D}+\sigma+s\chi)|\psi_n\rangle =
\langle \psi_n|\chi|\psi_n\rangle+\lambda_n \partial_s \langle \psi_n|
\psi_n\rangle=\langle \psi_n|\chi|\psi_n\rangle\,,
\end{eqnarray}
where we have used that $\langle
\partial_s\psi_n|(-\fssd{D}+\sigma)|\psi_n\rangle=\langle
\partial_s\psi_n|\psi_n\rangle\lambda_n$ and $\langle
\psi_n|(-\fssd{D}+\sigma)|\partial_s \psi_n\rangle=\lambda_n \langle
\psi_n|\partial_s \psi_n\rangle$. We arrive at
\begin{eqnarray}\label{eq:leading}
u_2(0)\chi^2 = \sum_{m_0<n_0}
\langle \psi_{n_0}|\chi|\psi_{n_0}\rangle
\langle \psi_{m_0}|\chi|\psi_{m_0}\rangle
\,\det{}' M_\psi\,.
\end{eqnarray}
Equation \eq{eq:leading} extends to general $u_n(0)\chi^n$ with $n\leq \Nf$. We
are specifically interested in $\Nf =3$ with the remaining cubic term
$u_3(\sigma)\sigma^{\Nf -3}\chi^3=u_3(\sigma)\chi^3$, in leading order,
\begin{eqnarray}\label{eq:leadingcub}
u_3(0)\chi^3 = \left(\prod_{n_0}\langle \psi_{n_0}|\chi|\psi_{n_0}\rangle
\right) \,\det{}' M_\psi\,.
\end{eqnarray}
This proves \eq{eq:instpot}. We can also directly use
Eqs.~\eqref{eq:leading},\eqref{eq:leadingcub} to compute the $\chi^2$
and $\chi^3$ coefficients as group averages $\langle u_2\rangle_{\rm
  SU(3)} $, $\langle u_3\rangle_{\rm SU(3)} $ with the help of
\cite{Creutz:mg}. We arrive at
\begin{eqnarray}\nonumber
\frac{3}{\chi^2}
\left\langle\prod_{n_0=1}^2\langle \psi_{n_0}|\chi|\psi_{n_0}\rangle \right
\rangle_{\rm SU(3)}&=& \frac{1}{72}
\,,\\
\frac{1}{\chi^3}
\left\langle\prod_{n_0=1}^3\langle \psi_{n_0}|\chi|\psi_{n_0}\rangle \right
\rangle_{\rm SU(3)}&=& \frac{1}{648\sqrt{6}}
\,,
\label{eq:group} \end{eqnarray}
leading to \eq{potential}.

\section{IR running coupling effects in the lowest-order
  approximation}
\label{runcoup}

\begin{figure}[t]
\begin{center}
\subfigure[]{\scalebox{0.65}[0.65]{
\begin{picture}(190,140)(40,0)
\includegraphics[width=9.5cm]{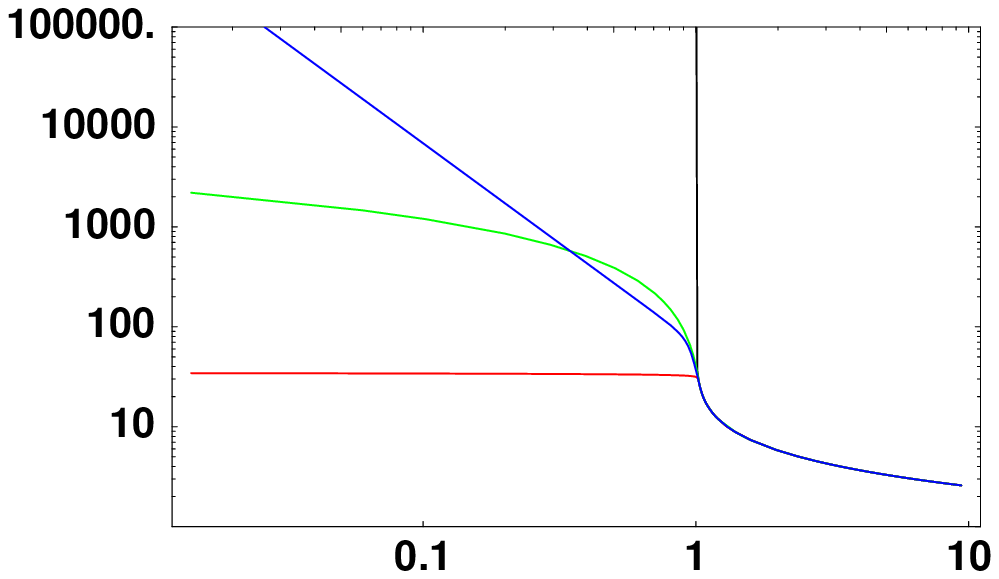}
\Text(-10,-05)[c]{\scalebox{1.6}[1.6]{${1}/{\rho}$}}
\Text(-300,150)[c]{\scalebox{1.6}[1.6]{$g^2(\frac{1}{\rho})$}}
\end{picture}
}}
\hspace{3.5cm}
\subfigure[]{\scalebox{0.65}[0.65]{
\begin{picture}(190,140)(20,0)
\includegraphics[width=9.5cm]{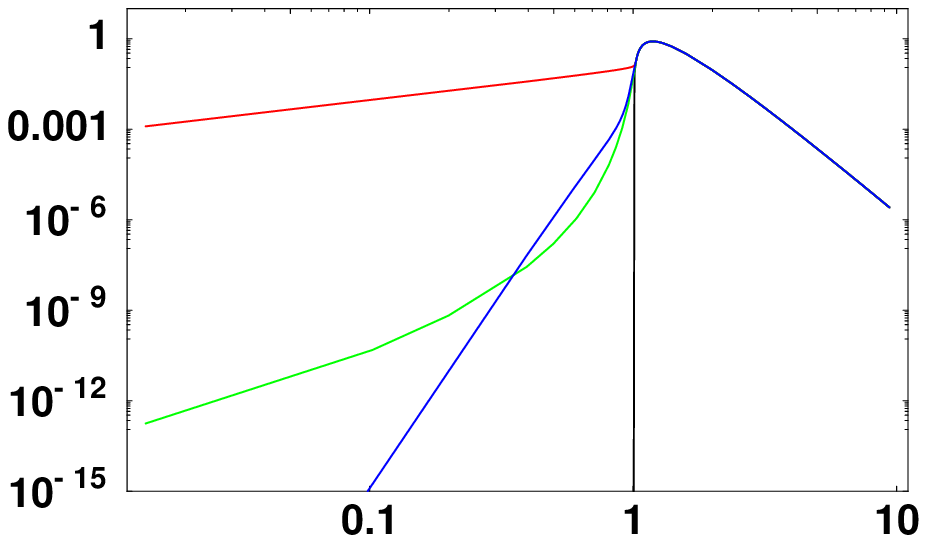}
\Text(-10,-05)[c]{\scalebox{1.6}[1.6]{${1}/{\rho}$}}
\Text(-290,150)[c]{\scalebox{1.6}[1.6]{$\rho
\frac{dn_{\textrm{inst}}}{d\rho}$}}
\end{picture}
}}
\end{center}
\vspace{-1.0cm}
\caption{Instanton density (right panel) for various types of infrared
  behavior for the gauge coupling (left panel). Red: $g\sim
  \textrm{const.}$, green: $g \sim \ln(\rho)$, blue: $g\sim \rho^2$,
  black: one loop behavior with $g=\infty$ for
  $\frac{1}{\rho}<\Lambda_{\textrm{QCD}}$. Strong non-perturbative
  behavior is modeled to set in at roughly $g^2\approx30$.  We note
  that the integral over the instanton size remains finite. More
  quantitatively, this holds if the plotted quantity on the right
  panel vanishes faster than $\frac{1}{\rho^{\epsilon}}$ with
  $\epsilon>0$ in the infrared. Most importantly, the constant $\zeta$
  occurring in our lowest order approximation to the effective
  potential, Eq.~\eqref{potential}, remains a finite number.}
\label{density}
\end{figure}

Here we demonstrate that the qualitative features of the
instanton-induced effective potential to lowest order in the scalar
condensates is largely independent of the behavior of the running
coupling. This can be deduced from a study of the $\rho$ integration
in Eq.~\eqref{potential}.  For constant $g^2(\rho)=g^2$ and $\Nf<4$,
the $\rho$ integral is infrared ($\rho\to\infty$) convergent, but has
a (naive) UV divergence. The one-loop running removes this UV
divergence, because of asymptotic freedom. Then, the integration
kernel behaves as
$\sim\rho^{\frac{11}{3}\Nc+\frac{1}{3}\Nf-5}(\ln(\rho))^{2\Nc}$ for
small $\rho$, rendering the integral convergent in this regime. In the
infrared, the situation is less clear, since one-loop running is
certainly not a valid approximation for the gauge coupling.
Nevertheless, the restrictions on the behavior of $g^2(\rho)$ for
$\rho\to\infty$ are rather mild. Indeed, for positive and well-defined
$g^2(\rho)$ there are no restrictions at all for $\Nf<4$ (massless
flavors). We can even allow for a diverging coupling at a finite
infrared scale $\rho_{\text{div}}$. In this case, it is reasonable to
assume that the coupling remains infinite for even larger distance
scales, such that the integrand remains exactly zero for all
$\rho>\rho_{\text{div}}$.  In Fig.~\ref{density} we plot the integrand
for running couplings with different infrared behavior. It is our main
conclusion that all reasonable forms for the running coupling in the
infrared imply a finite constant $\zeta$ in Eq.~\eqref{potential}.

\end{appendix}

\end{document}